\documentclass[conference]{IEEEtran}
\usepackage{amsmath,amssymb,amsfonts}
\IEEEoverridecommandlockouts

\newcommand{\new}[1]{#1}

\newcommand{\ignore}[1]{}

\usepackage{balance}  \usepackage{url}
\usepackage{microtype}
\usepackage{multirow}
\usepackage{pgfplotstable}
\usepackage{afterpage}
\usepackage{placeins}

\usepackage[mathcal]{euscript}

\usepackage{soul}
\usepackage{algorithm}
\usepackage{algpseudocode}
\usepackage{pgfplots}
\usepackage{color}
\usepackage{subcaption}
\usepackage{booktabs}
\usepackage{dcolumn}
\usepackage{tabularx}
\usepackage{tikz}
\usepackage[normalem]{ulem}

\usetikzlibrary{positioning}
\pgfplotsset{compat=1.7}
\usetikzlibrary{patterns}
\usepgfplotslibrary{groupplots}
\usepackage{pgfplots,pgfplotstable}
\usepgfplotslibrary{dateplot}
\usetikzlibrary{pgfplots.groupplots}
\usepackage{tablefootnote}

\newtheorem{definition}{Definition}

\newlength{\oldtextfloatsep}
\newcolumntype{L}[1]{>{\raggedright\let\newline\\\arraybackslash\hspace{0pt}}m{#1}}
\newcolumntype{C}[1]{>{\centering\let\newline\\\arraybackslash\hspace{0pt}}m{#1}}
\newcolumntype{R}[1]{>{\raggedleft\let\newline\\\arraybackslash\hspace{0pt}}m{#1}}

\definecolor{RYB1}{RGB}{141, 211, 199}
\definecolor{RYB2}{RGB}{255, 255, 179}
\definecolor{RYB3}{RGB}{190, 186, 218}
\definecolor{RYB4}{RGB}{251, 128, 114}
\definecolor{RYB5}{RGB}{108, 157, 201}
\definecolor{RYB6}{RGB}{253, 180, 98}
\definecolor{RYB7}{RGB}{179, 222, 105}

\definecolor{PYBLUE}{RGB}{70,118,178}

\definecolor{plot1}{RGB}{0,107,164}
\definecolor{plot2}{RGB}{255,128,14}
\definecolor{plot3}{RGB}{51,160,44}
\definecolor{plot4}{RGB}{178,223,138}
\definecolor{plot5}{RGB}{117,112,179}
\definecolor{plot6}{RGB}{217,95,2}
\definecolor{plotGray}{RGB}{89,89,89}

\pgfplotscreateplotcyclelist{colorbrewer-RYB}{
{RYB1!50!black,fill=RYB1},
{RYB2!50!black,fill=RYB2},
{RYB3!50!black,fill=RYB3},
{RYB4!50!black,fill=RYB4},
{RYB5!50!black,fill=RYB5},
{RYB6!50!black,fill=RYB6},
{RYB7!50!black,fill=RYB7},
}

\newcommand{\eg}{e.g.}

\newcommand{\system}{\textsc{Tahoma}}

\newcommand{\cmNoCost}{\textsc{Infer Only}}
\newcommand{\cmLoad}{\textsc{Ongoing}}
\newcommand{\cmResize}{\textsc{Camera}}
\newcommand{\cmLoadResize}{\textsc{Archive}}

\newcommand{\minisection}[1]{\vspace{0.1cm} \noindent {\bf #1} ---}

\begin{document}
\title{Physical Representation-based Predicate Optimization for a Visual Analytics Database}

\author{\IEEEauthorblockN{Michael R. Anderson}
\IEEEauthorblockA{\textit{University of Michigan}\\
mrander@umich.edu}
\and
\IEEEauthorblockN{Michael Cafarella}
\IEEEauthorblockA{\textit{University of Michigan}\\
michjc@umich.edu}
\and
\IEEEauthorblockN{German Ros$^*$\thanks{$^*$Work done while at Toyota Research Institute.}}
\IEEEauthorblockA{\textit{Intel Labs}\\
german.ros@intel.com}
\and
\IEEEauthorblockN{Thomas F. Wenisch}
\IEEEauthorblockA{\textit{University of Michigan}\\
twenisch@umich.edu}
}

\maketitle

\begin{abstract}
Querying the content of images and video requires expensive content extraction methods. Modern extraction techniques are based on deep convolutional neural networks (CNNs) and can classify objects within images with astounding accuracy.   Unfortunately, these methods are slow: processing a single image can take about 10 milliseconds on modern GPU-based hardware. As massive video libraries become ubiquitous, running a content-based query over millions of video frames is prohibitive.

One promising approach to reduce the runtime cost of queries of visual content is to use a hierarchical model, such as a cascade, where simple cases are handled by an inexpensive classifier.  Prior work has sought to design cascades that optimize the computational cost of inference by, for example, using smaller CNNs.  However, we observe that there are critical factors besides the inference time that dramatically impact the overall query time.  Notably, by treating the physical representation of the input image as part of our query optimization---that is, by including image transformations such as resolution scaling or color-depth reduction within the cascade---we can optimize data handling costs and enable drastically more efficient classifier cascades.

In this paper, we propose \system, which generates and evaluates many potential classifier cascades that jointly optimize the CNN architecture and input data representation. Our experiments on a subset of ImageNet show that \system's input transformations speed up cascades by up to 35 times.  We also find up to a 98x speedup over the ResNet50 classifier with no loss in accuracy and a 280x speedup if some accuracy is sacrificed.
 \end{abstract}

\section{Introduction}

Recent developments in computer vision have made feasible a long-term dream for the database community: a {\em visual analytics database}, which stores image data and answers user questions about its contents.  For example, video frames from a city's traffic cameras could be used to count cars per minute.  Or photos uploaded to photo storage web sites could be automatically sorted and tagged based on their contents. The sheer volume and diversity of data captured by cameras opens up myriad analytical query possibilities, if the content hidden behind opaque pixel values can be extracted at scale.

Deep convolutional neural networks (CNNs)---the family of methods used in modern computer vision systems---have enabled huge strides in image understanding in the last few years through tasks like image classification and object detection. For example, the classification error rate in the annual ImageNet Large-Scale Visual Recognition Challenge (ILSVRC)~\cite{ILSVRC15} dropped from 25\% to 18\% in 2012 when a deep CNN was first used~\cite{krizhevsky2012imagenet}. Recent results have lowered the error rate to 2\%~\cite{hu2017squeeze}, rivaling or exceeding human performance.

Unfortunately, deep networks pose a considerable computational challenge when deployed in an analytical database system: a model's inference for a single image can require a lengthy series of large tensor multiplications. For example, YOLOv2, an object detection system designed for speed, requires 8.52 billion operations per single 416x416 pixel image, processing about 67 images per second on a modern GPU~\cite{redmon2016yolo9000}.  Since GPU hardware is far more expensive than most image sensors, data from multi-camera applications will soon outpace processing capabilities.  Simply, to query huge amounts of image data, we need drastically lower processing costs.

Processing queries over a corpus of image data fits a more general \emph{loop-and-test} pattern that is common to many machine learning tasks: the processor loops over the data, executing an expensive operator on each element to find those satisfying the task's constraints. In this case, content is extracted from each image by the expensive inference stage of a deep network to determine if the image satisfies a binary predicate specified in a user's query. While a loop-and-test process can be shortened by processing fewer items overall---using simple sampling or more sophisticated input selection~\cite{anderson2016input}---we focus here on speeding up the \emph{test} phase by reducing the per-image inference cost.

Recent work has reduced inference times for these types of deep learning systems (e.g., ~\cite{hinton2015distilling,kang2017noscope}). However, we note that all of the visual data system optimizations to date suffer from a critical defect: they concentrate only on computation and ignore the inevitable data-handling costs, such as loading and transformation.  Any query optimization method that focuses only on reducing computational load cannot exploit complex {\em data-centric tradeoffs} that weigh the amount of image data available, the classifier accuracy, and data handling costs.

As an example, consider an optimizer choosing between two image classification models ($M_1$ and $M_2$). $M_1$ accepts a 3-channel, full-color, 224x224 image as input while $M_2$ accepts a 1-channel, grayscale, 224x224 image. $M_1$ has fewer convolutional layers and, despite the larger input, requires fewer tensor operations than $M_2$, so its inference is faster.  $M_2$ uses less rich data than $M_1$, but is nonetheless able to obtain comparable accuracy because of its additional convolutional layers.  An optimizer that considers only the inference speed would choose $M_1$. However, a data system using $M_2$ might be faster, as its inputs load in one-third the time of those of $M_1$.

Such data-handling tradeoffs are important because visual analytics systems are likely to be architecturally diverse. Some systems may store multiple versions of the same image data (high-res vs. low-res). Others may employ different storage systems (local server vs. cloud vs. in-camera).
In some architectures---say, one in which connectivity conditions change, or one in which GPUs are only intermittently available---the highest-payoff query plan may change by the moment.  Ignoring data-centric tradeoffs can sacrifices substantial performance. In this work, we propose a framework for handling data-centric tradeoffs when optimizing visual analytical queries.

\minisection{Technical challenge}  We aim to optimize visual analytics queries that might be run in a range of diverse architectures and deployment settings.  In this work, we ignore many standard relational query optimization issues, such as query plan rewriting or indexing. We focus exclusively on designing and choosing the CNN-based operator that implements an image-sensitive relational predicate. These operators can be chosen in a manner that trades system throughput against classification accuracy.  Making that tradeoff of runtime vs. data quality is an application-specific decision we leave to the user.  This paper offers a framework for identifying the best possible operator implementation, subject to a user's desired tradeoff.

\minisection{Our approach}
 One method to speed up an expensive-but-accurate CNN is to replace it with cascades of fast, high precision (but low recall) image classifiers~\cite{kang2017noscope,cai2015learning,sun2013deep}. This is effective but focuses exclusively on computational efficiency.  We start similarly, training a large number of specialized candidate binary-classification CNN models by varying not only CNN hyperparameters (as in prior work~\cite{kang2017noscope}), but also the representations of the inputs; for example, we build an $n$-layer CNN for large full-color inputs, another for small full-color inputs, more for grayscale inputs, and so on.

 From these core candidate models, we then construct a massive number of classifier cascades.  All of these cascades have different initially unknown runtime and accuracy characteristics.  Our optimization method efficiently evaluates the cascades' accuracy using held-out data, and evaluates their runtime characteristics for the system's current deployment scenario.  Finally, it identifies the Pareto-optimal cascades that satisfy the user's application-specific speed and accuracy constraints.

\minisection{Organization}
After formally describing our problem in Section~\ref{sec:problemStatement}  we discuss the following contributions:

\begin{itemize}
    \item We propose a method for identifying high-quality image predicate implementations, by exploring CNN hyperparameters and varying input data representation (Section~\ref{sec:cascades}).
    \item We show the dramatic impact on runtime when a system is running in different deployment scenarios and is aware of the deployment-specific data handling costs (Section~\ref{sec:costs}).
    \item We prototype our methods in a system called \system\ and show that it provides up to a 35x speedup to classifier cascades through input data transformations. \system\ also achieves up to a 98x speedup over the ResNet50 image classifier with no accuracy loss (Section~\ref{sec:experiments}).
\end{itemize}

We follow with a discussion of related work in Section~\ref{sec:relatedWork}.
 
\section{Background}
\label{sec:background}

In this section, we will briefly give some background on cascade classifiers, since these methods are core building blocks of our work.
Numerous systems have used cascades to speed up classifier inference. One of the first was the Viola-Jones face detector~\cite{viola2004robust}, which used a series of  classifiers based on simple image features to detect faces in subsections of photographs; if any classifier had high confidence in its result, the result was immediately accepted, avoiding the use of further classifiers.

\begin{figure}
    \centering
    \includegraphics[width=.85\linewidth]{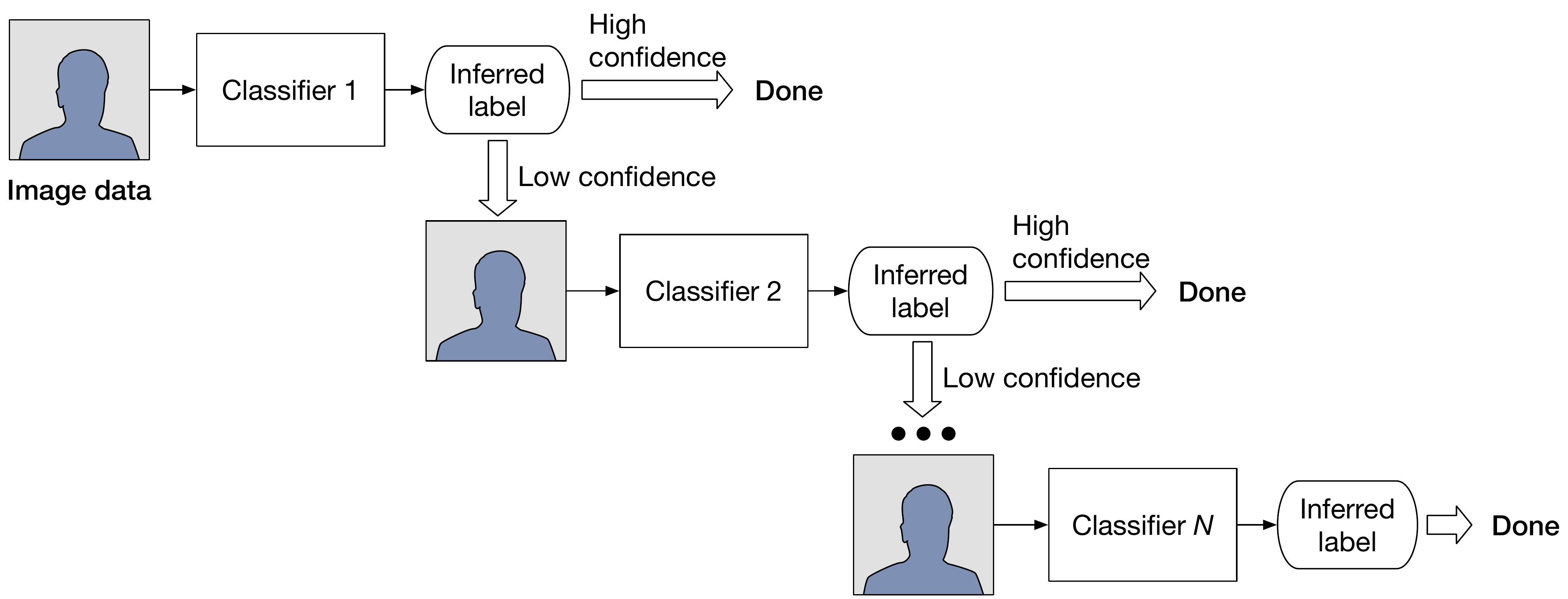}
    \caption{A multi-level classifier cascade. If the first classifier's output is uncertain, the input is classified by the second, etc. If reached, the final classifier's output is accepted as the label.}
    \label{fig:cascadeDiagram}
    \vspace{-0.2cm}
\end{figure}

Figure~\ref{fig:cascadeDiagram} illustrates the general cascade process: an image is input into a classifier, whose output is accepted if it has high confidence. Otherwise, a second classifier is used, where again, a low-confidence result will send the input to a third, and so on. If the final level of the cascade is reached, its output is accepted. Ideally, the initial levels of a cascade are fast with high precision, though they may suffer from low recall. In face detection, for example, if most images have no human faces at all, the first classifier may quickly eliminate most cases.  Only the few cases containing a face will be processed by the remainder of the cascade, at correspondingly higher cost.

Recently, cascades have been used to accelerate the relatively slow inference speeds of deep neural networks~\cite{kang2017noscope,cai2015learning,sun2013deep}. Our system takes these techniques further by exploiting the representations of the inputs and finding optimal cascades for a user's deployment scenario.   \section{Design Considerations}
\label{sec:designConsiderations}
Several key questions are important to \system's design. We touch on these issues below.

\minisection{Issue 1: Object detection vs. image classification}
In general terms, an object detector (e.g., YOLOv2~\cite{redmon2016yolo9000}) finds the location of particular object classes within an image, while an image classifier (e.g., ResNet50~\cite{he2016deep}) identifies the overall contents of an image as one of a particular set of classes. When implementing binary \texttt{contains-object} predicates in a visual database application, the extra architectural and computational complexity of object detectors is unnecessary. Our cascades therefore use small and fast CNN image classifiers. Applications requiring object location within images, however, could build cascades from small and fast object detectors.

\minisection{Issue 2: Online vs. offline classification}
If a user's corpus is small or slow-growing enough to allow for offline classification of the entire dataset with available resources, this paper's techniques are unnecessary. Also, our techniques may be unneeded if query predicates are fixed and new ones are unlikely to be introduced.  In such cases, the user could materialize the classification results for each image on ingest and store them in a standard database for future queries.

However, there are many exciting real-world applications where assumptions of a small or slowly growing dataset or stable query predicates are unrealistic. Consider, for example, a website for photo storage: Flickr has reported that days of heavy usage can see 25 million photo uploads~\cite{flickrblog}. Facebook had 350 million daily photo uploads as of 2013~\cite{facebookwhitepaper}. Video applications can be even more extreme. A single self-driving car can have over a dozen cameras, each gathering 30 frames per second for many hours on a daily basis; a fleet of such cars can generate a huge amount of data.

Further, many applications require retrospective exploration and analysis, where the query predicates may not be known in advance. For example, a self-driving car engineer may wish to find historical examples of a new failure case, requiring the training of a new model to be run over an existing corpus. Or consider a police investigation reviewing thousands of hours of surveillance camera footage to find a local delivery van with a unique logo: a trained object detector would be ideal, but such a specific model is unlikely to already exist. In general, it is unlikely that all possible query predicates can be enumerated in advance for a visual analytics application.

\minisection{Issue 3: Training costs}
A deep CNN image classifier, such as ResNet50~\cite{he2016deep}, can take days  to fully train, due to the model's architectural complexity and the huge training set needed for such a complex model. Such a burden is incompatible with our use case, since requiring this amount of time to install a new predicate in our system is unreasonable. Thankfully, our simple binary predicates typically do not require huge classifiers. The majority of our cascades consist solely of small, specialized classifiers, which train in just minutes.

Further, in cases where a deep network may be needed, training it from scratch is likely to be unnecessary. It is common practice to fine-tune pre-trained deep networks to a particular task~\cite{jia2014caffe}. Generally, when fine tuning, most of the deep network is frozen and only the last several layers are modified and retrained to a new task, taking advantage of already-learned features from a similar problem domain. In our experiments, we fine tuned ResNet50 for binary classification tasks using a modern GPU in only 2 to 4 hours.

\minisection{Issue 4: Deployment scenarios}
A visual analytics system may be deployed in a variety of scenarios, requiring accounting for differing data handling costs, in addition to classifier inference costs. Consider the following example scenarios, also later used in our experiments:

\begin{itemize}
    \item \cmLoadResize\ -- In this situation, a large archival corpus of  historical image data is stored on local drives. Each image must first be loaded from the drive and then transformed into an appropriate input format for the classifier.

    \item \cmLoad\ -- Here, video is continually ingested from its source into a datacenter-based query system, where it is transformed into appropriate representations that are stored on SSD for later queries. Because this data is transformed as it is acquired, only the cost of loading the representations from disk are considered at query time.

    \item \cmResize\ -- If compute nodes are at the edge of the network (e.g., connected to surveillance cameras), the images can be directly provided to the classifiers. Only the image transformation costs must be considered, since transfer costs from camera to memory are negligible.
\end{itemize}

We show in Section~\ref{sec:costModels} that scenario-awareness while evaluating of the accuracy and speed of classification systems can lead to a large practical increase in a system's throughput.
 \section{Definitions and Notation}
\label{sec:problemStatement}
This section formalizes the particular image classification problem addressed here. Table~\ref{tab:notation} lists frequently used notation.

\minisection{Content-based Queries}
A query system that operates over images can perform queries over two main types of information:  image metadata (e.g., GPS or acceleration information that may accompany frames from dashcams) and content extracted from images themselves (e.g., the image contains a bicycle). Metadata queries are easily handled by existing methods, so we focus this work on processing \textit{content-based queries}.

\begin{definition}[Content-based query]
Given a corpus of image data $\mathcal{I}$ containing images $I_1,\ldots,I_n$, the tuples $T_i = (t_{i1},\ldots,t_{im})$ represent the visual contents for each image $I_i$, where each element $t_{ij}$ in this tuple represents a content object present in image $I_i$. A content-based query is constructed of predicates that can be evaluated with the elements of $T_i$.
\end{definition}

We restrict content-based queries to \textit{binary queries}, which can be combined into a full query over all $T_\mathcal{I}$ tuples.

\begin{table}
    \small
    \centering
    \caption{Frequently used notation. }
    \begin{tabular}{ll}
    \toprule
         Notation & Definition \\
    \midrule
         $\mathcal{I} = (I_1,\ldots,I_n) $ & Image data corpus \\
         $T_i = (t_{i1},\ldots,t_{in})$ & Content tuple for image $I_i$ \\
         $K$ & Image classifier, s.t. $K(I_i) = t_ij$ \\
         $\mathcal{M} = (M_1,\ldots,M_m)$ & Basic classification models \\
         $\mathcal{A} = (A_1,\ldots,A_{n_a})$ & Model architecture specifications \\
         $\mathcal{F} = (F_1,\ldots,F_{n_f})$ & Input transformation functions \\
         $C = (M_1,\ldots,M_n)$ & Cascade of $n$ models belonging to $\mathcal{M}$\\
         $\mathcal{C} = (C_1,\ldots,C_k)$ & Collection of cascades for a binary query\\
         $p_\text{low}, p_\text{high}$ & Cascade model decision thresholds\\
    \bottomrule
    \end{tabular}
    \label{tab:notation}
\end{table}

\begin{definition}[Binary query]
Given an image's content tuple $T_i$, a binary query involves a single \texttt{contains-object} predicate evaluated with a single tuple element, $t_{ij}$. A binary query thus asks if image $I_i$ contains object $t_{ij}$.
\end{definition}

Despite this restriction, our envisioned system will support complicated queries that can be rewritten as a combinations of metadata and binary query predicates. For example, the query ``Find images from Detroit containing a bicycle'' can be decomposed into a metadata predicate (location = `Detroit') and a binary query predicate (\texttt{contains\_object}(bicycle)). Our focus is on choosing the best classifiers to implement \new{a given \texttt{contains\_object}} predicate. \new{While further query optimization could be done considering multiple binary predicates in concert, we leave that for future work and here concentrate on optimizing single predicates. }

\minisection{Image Classification}
The content tuples are generally not available upon image acquisition and must be extracted from the image. In this work, we focus on extraction via \textit{image classification}. In particular, we focus on queries where classification has not been performed in advance, so that classification must be performed as part of query execution.

\begin{definition}[Image classification]
Given a corpus of image data $\mathcal{I}$, for each image $I_i \in \mathcal{I}$, we wish to generate a binary label $L$ using a classifier $K$, such that $L_i = K(I_i)$. The label $L_i$ corresponds to a member $t_ij$ of content tuple $T_i$.
\end{definition}

The output of a classifier model can be thought of as a virtual column in a relation describing the content objects in images. For example, processing the query predicate \texttt{contains\_object}(bicycle) would populate the bicycle column of this relation with the output of $K_\text{bicycle}$.
The classifier $K$ could be generated by a range of methods, including \textit{basic models} like logistic regression or deep CNNs, or may be a collection of basic models, such as \textit{classifier cascades}.

\begin{definition}[Basic model]
A basic model $M$ implements a classification method that accepts image $I$ as input and outputs a binary classification result. Our system generates a large set of such CNN-based models $\mathcal{M} = (M_1,...,M_m)$.
\end{definition}

Two factors parameterize a model $M$: \textit{model architecture specification} $A$ and \textit{input transformation function} $F$.

\begin{definition}[Model architecture specification]
The internal architecture of a model $M$ is specified by $A$. For the CNNs used in our system, $A_m$ describes network hyperparameters, such as the number and size of layers.
$\mathcal{A} = (A_1,\ldots,A_{n_a})$ gives all potential architectural options for models in $\mathcal{M}$.
\end{definition}

\begin{definition}[Input transformations]
Before the classification of an image $I$ by model $M$, the raw image data is processed by an image transformation function $F$, such that input image $I$ is transformed into output image $I'$. Such a function may perform one or more operations such as resizing, normalizing, or reducing color depth.   The set $\mathcal{F} = (F_1,\ldots,F_{n_f})$ gives all functions available to pre-process image data for models in $\mathcal{M}$.
\end{definition}

We use the cross product of  $\mathcal{F}$ and $\mathcal{A}$ as the model design space, resulting, in practice, in several hundred individual models in $\mathcal{M}$ for each binary query. A goal of this work is to determine which models are most suitable for a user's accuracy and runtime constraints and current deployment scenario.

\minisection{Classifier cascades}
Classification models can be aggregated into collections or ensembles to improve either accuracy or speed. One such method designed to improve classification speed is the \textit{classifier cascade}~\cite{viola2004robust}.

\begin{definition}[Classifier cascade]
A classifier cascade $C = (M_1,\ldots,M_n)$ is a list of $n$ basic models with probabilistically interpretable output. The models are run in series: image $I_i$ is classified by $M_1$, and if the output is between two given decision thresholds, $p_{\text{low}}$ and $p_{\text{high}}$, it is \textit{uncertain}. If so, $I_i$ is then classified by $M_2$. Otherwise, the cascade is stopped and $M_1$'s output is accepted as the label of $I_i$. This continues to the final classifier $M_{n}$, whose output is always accepted.
\end{definition}

Given a set of classification models $\mathcal{M}$, we can construct a large set of cascades $\mathcal{C} = (C_1,\ldots,C_k)$ with up to $n$ levels each, to be evaluated in terms of accuracy and throughput.

\minisection{Model evaluation}
The quality of classifier---either a model $M$ or a cascade $K$---is given by its \textit{accuracy} and its \textit{throughput}. Accuracy gives the fraction of labels produced by $M$ that are correct.  Throughput is the number of classifications per unit time and measures how fast the model's relation is populated.

For a set of classifiers, we can find a subset that is Pareto-optimal over these two criteria. That is, there is a subset that is non-dominated in terms of accuracy and throughput, from which users can select a classifier to meet application needs.

\new{
\minisection{Problem statement}
With the preceding definitions, we can formally describe the problem addressed in this paper thusly:

\vspace{0.2cm}
\emph{For an image corpus $\mathcal{I}$ and a set of binary classification models $\mathcal{M} = (M_1,\ldots,M_m)$ parameterized by architectural specifications $\mathcal{A} = (A_1,\ldots,A_{n_a})$ and input transformation functions $\mathcal{F} = (F_1,\ldots,F_{n_f})$, find the set of classifier cascades $\mathcal{C} = (C_1,\ldots,C_k)$ constructed from models in $\mathcal{M}$ that are Pareto-optimal in terms of accuracy and throughput over $\mathcal{I}$.}
}
 
\section{Cascade Methodology}
\label{sec:cascades}

 In this section, we give an overview of system architecture and discuss the details of our models and cascades.

\begin{figure}
    \centering
    \includegraphics[width=.8\linewidth]{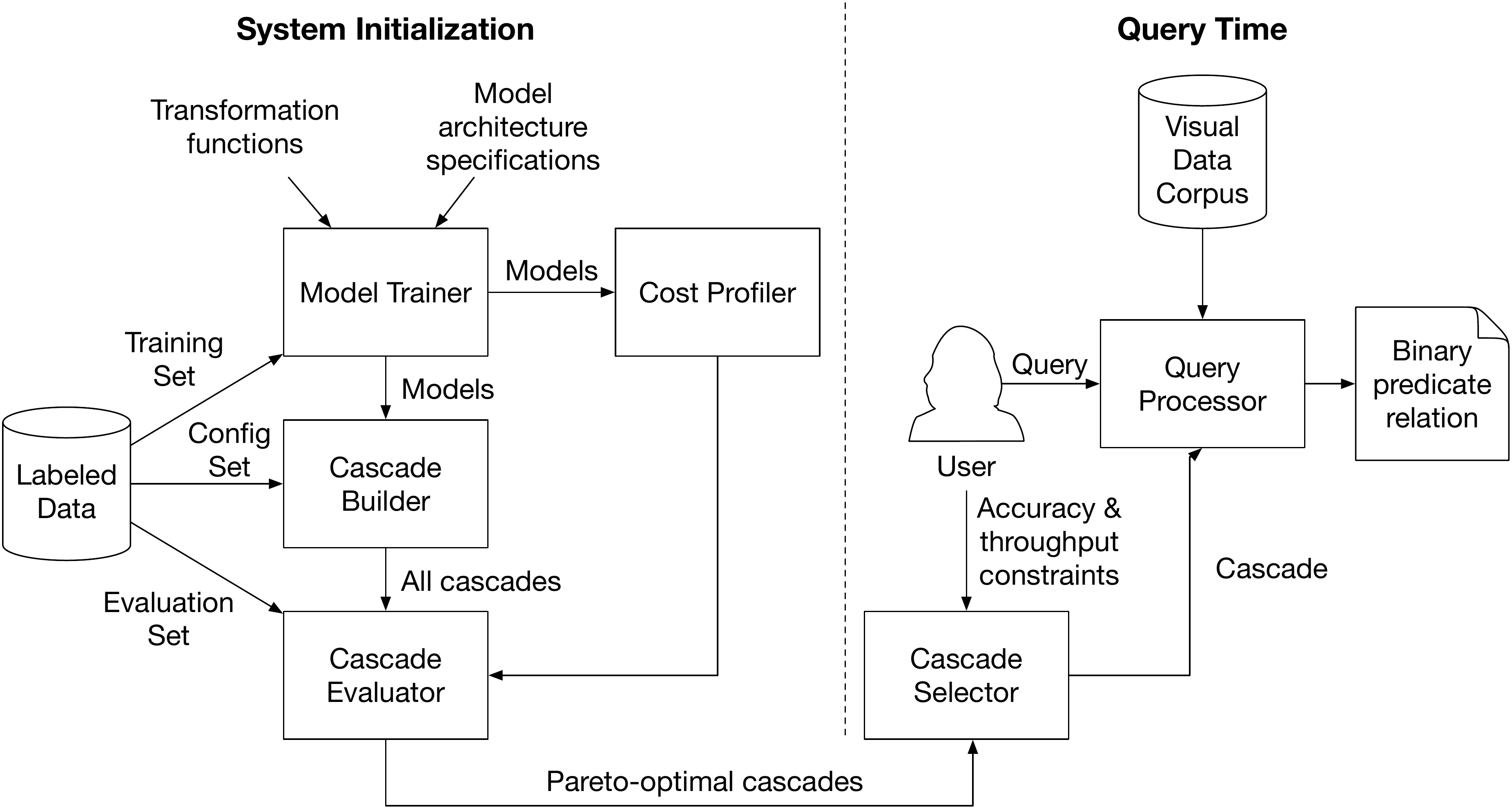}
    \caption{\system\ architecture}
    \label{fig:arch}
    \vspace{-.2cm}
\end{figure}

\subsection{System Architecture}

Figure~\ref{fig:arch} sketches out \system's architecture. \system\ has two main modes of operation: system initialization and query execution. During system initialization, model repository is prepared for each binary predicate, which requires a set of labeled data. This dataset is small compared to what is generally used to train deep CNNs: per binary predicate, \system\ requires 3,000--4,000 labeled images, with equal numbers of positive and negative examples. The labeled data is split into three sets for training, configuration, and evaluation. Training set $\mathcal{I_\text{train}}$, transformation functions $\mathcal{F}$, and architecture specifications $\mathcal{A}$ are provided as input to the model trainer.

For a given binary predicate, a set of models $M$ (each implementing the \texttt{contains\_object} operator) is trained and provided to the cost profiler and to the cascade builder.  The cost profiler measures the throughput of each model in the current deployment scenario (see Section~\ref{sec:costs}). The cascade builder constructs \system's cascade set $\mathcal{C}$, using all possible combinations of size $n$ of the models in $\mathcal{M}$ and the configuration set $\mathcal{I_\text{config}}$ (see Section~\ref{sec:decisionThresholds}). Using the evaluation set $\mathcal{I_\text{eval}}$, the cascade evaluator measures each cascade's accuracy and throughput (see Section~\ref{sec:evaluatingCascades}). With this, the system determines the set of Pareto-optimal cascades for use at query time.

Similar to how approximate query systems like BlinkDB~\cite{agarwal2013blinkdb} and VerdictDB~\cite{park2018verdictdb} allow users to specify approximation constraints as part of their queries, a \system\ user provides their constraints on accuracy ($U_\text{acc}$) and throughput ($U_\text{thru}$) at query time
\new{(in the form of the highest tolerable loss in either of those parameters).} The cascade selector chooses which of the Pareto-optimal cascades best suits the user's desired tradeoff. \new{For example, the user may wish to maximize throughput as long as the resulting cascade does not suffer more than a 5\% loss in accuracy over the most accurate cascade available. The user would set $U_\text{acc} = 0.05$ and provide no constraint for $U_\text{thru}$. The system would select the cascade from the set of Pareto-optimal cascades has an accuracy closest to (but not below) 95\% of that of the most accurate cascade. Because this is a Pareto-optimal choice, there will be no faster cascades at that (or a higher) accuracy level.} The selected cascade processes the data in the corpus, extracting the notional relation for the binary predicate in the user's query.

\new{
\minisection{Integration considerations}
Because our goal was to explore the optimization methods discussed in this paper, we implemented \system\ as a standalone query system. However, we believe future deployments of \system's ideas will likely be embodied in RDBMS software. The execution of a \texttt{contains\_object} operator is analogous to that of a user-defined function (UDF) in a database such as PostgreSQL, and could be wrapped in the RDBMS \texttt{CREATE FUNCTION} statement. RDBMS query optimizers could leverage additional metadata relations, such as image location and capture date, to reduce the number of expensive \system\ UDFs calls for a specific query. Further, UDF output could be stored as a partially materialized table, enabling further query optimization.

\system's initialization process is run at the installation of each new binary predicate. During this, the profiled speeds could be used to inform the RDBMS query optimizer of the execution cost of the UDF (like the PostgreSQL's \texttt{COST} parameter for \texttt{CREATE FUNCTION}, if each Pareto-optimal cascade was implemented as a separate UDF). While indexing or materializing the output of the UDFs at system installation is not practical in our envisioned deployment scenarios, database triggers could be used to execute the \system\ UDFs over newly ingested data after system initialization to pre-materialize the output for future queries. In such situations, slower processing may be tolerated for more accurate results, allowing a different Pareto-optimal cascade choice than at query time.
}

\subsection{Building Models}
With \system, we create a huge number of cascades $\mathcal{C}$ by first training a large number of individual classification models $\mathcal{M}$. We build the collection of models in two ways: by varying the internal architecture of our CNN-based classifiers with our model architecture specifications $\mathcal{A}$ and by transforming the input images using the input transformation functions $\mathcal{F}$.

\begin{figure}
    \centering
    \includegraphics[width=\linewidth]{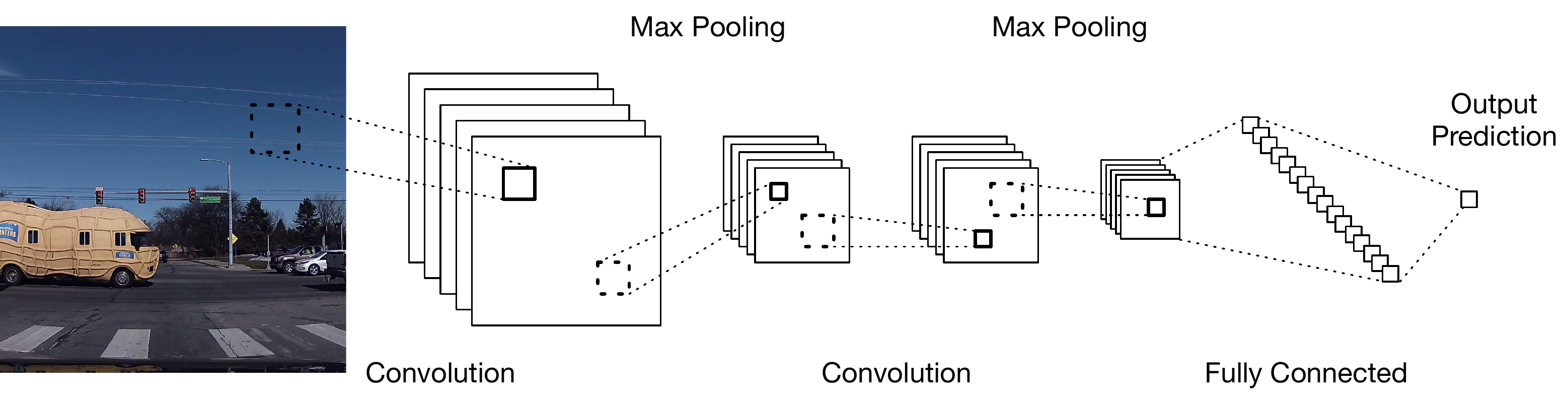}
    \caption{CNN architecture used by \system. The number of layers, as well as the number of nodes in each layer is varied as part of the model architecture specifications $\mathcal{A}$.}
    \vspace{-.2cm}
    \label{fig:cnn}
\end{figure}

\minisection{Model architecture variations}
\system\ uses convolutional neural networks for its models. Goodfellow et al. provide in-depth discussion on CNNs and deep learning~\cite{Goodfellow2016}. Our CNNs follow the basic architectural pattern shown in Figure~\ref{fig:cnn}. Input values are fed into one or more layers of convolutional nodes. Each convolutional layer is followed by a max pooling layer, connected by rectified linear activations (ReLu). The final convolutional layer feeds into a fully connected ReLu layer. A sigmoid output node provides the inferred label. A key point is that our CNNs are small (and thus fast), typically having only one to four convolutional layers. When creating models, we vary these architectural details according to $\mathcal{A}$; for our experimental settings of $\mathcal{A}$, see Section~\ref{sec:expCascadeConfigurations}.

\minisection{Input transformations}
We also vary the physical representation of the input to each model. The set of input transformation functions $\mathcal{F}$ comprises functions that perform one or more image processing operations, such as resolution scaling and color channel modification. These types of transformations are useful for building fast, small models: reducing image size and color depth can greatly reduce the number of model input values, directly reducing the size of the CNN's tensor operations. For our experiments, we scaled the image resolution (30x30, 60x60, 120x120, and 224x224 pixels), and for each image size, we used five different color variations (full 3-channel color, each of the individual red, green, and blue color channels, and single-channel grayscale).

\vspace{0.1cm}
The design space defined by these model architecture variations and input transformation functions result in hundreds of different model configurations (360 in our experiments). Once each model is trained on a labeled subset of $\mathcal{I}$, we can compose the models into cascades. Training can take less than a minute for the smallest networks (one convolutional layer with few nodes) with the smallest inputs (30x30 pixels, 1 color channel) to nearly an hour for the largest. Overall, training 360 models for a single binary predicate requires about 12 hours when done serially on an NVIDIA Tesla K80 GPU. Training is parallelizable, so this cost can be greatly reduced in practice.

\subsection{Computing Decision Thresholds}
\label{sec:decisionThresholds}

Each model in $\mathcal{M}$ provides a probabilistic output for a binary classification problem, a real number ranging from 0 to 1. Each model has a pair of decision thresholds, $p_{\text{low}}$ and $p_\text{high}$,
  which determine whether the model's labelling decision should be trusted. If the output $o \leq p_\text{low}$ or $o \geq p_\text{high}$, the model's output is accepted as the output of the cascade. If $p_\text{low} < o < p_\text{high}$, then we consider the model's output to be uncertain and reclassify the image with the next model in the cascade.

 These thresholds are chosen on a per-model basis, such that the precision of classification results with $o \leq p_\text{low}$ or $o \geq p_\text{high}$ matches a predefined constraint while recall is maximized. \new{Using a small configuration dataset distinct from the training set, the thresholds are determined via a grid search that sweeps through the potential thresholds to find those that provide a precision value greater than or equal to the target precision and selects the thresholds from those that maximize recall.}

\subsection{Constructing Cascades}

Each model has its decision thresholds determined independently, rather than in the context of a specific cascade. This assumption of independence allows us to quickly instantiate and evaluate the millions of possible multi-level cascades that can be constructed from the models in $\mathcal{M}$.

To determine the accuracy and throughput of our cascades, we first classify a set of labeled images $\mathcal{I}_\text{eval}$ with each model in $\mathcal{M}$. (The images in $\mathcal{I}_\text{eval}$ are distinct from those for training and determining decision thresholds so that the resulting accuracy measurements are not the product of overfitting.)
Since the cascades in $\mathcal{C}$ comprise combinations of the models in $\mathcal{M}$, the above evaluation need only be done once per model (360 times in our experiments) and not once per cascade (1.3 million).

Ensuring the independence of both model evaluations and decision thresholds (as described in Section~\ref{sec:decisionThresholds}) enables extremely fast evaluation of cascades: our evaluation required just over one minute to determine the accuracy and throughput values for 1.3 million cascades. As such, once models have been trained and classified, the selection of a cascade can be part of query planning at query execution time and can thus incorporate query-specific performance criteria (e.g., which storage devices are providing input images).

Using the pre-computed classification results for $\mathcal{I}_\text{eval}$ for each model in the cascade, the cascade execution is simulated to obtain its  label predictions for $\mathcal{I}_\text{eval}$. The cascade's accuracy is then computed by comparing with true labels for $\mathcal{I}_\text{eval}$.

\subsection{Evaluating Cascade Sets}
\label{sec:evaluatingCascades}

Each cascade in $\mathcal{C}$ can be displayed on a plot of accuracy versus throughput, as shown in Figure~\ref{fig:paretoExample}. Those with the best tradeoffs between accuracy and throughput belong to the Pareto frontier. Points on the Pareto frontier are those not dominated by any other points~\cite{Papadimitriou:2001}. Computing the Pareto frontier over two attributes, as we do, is $O(n\log n)$ in the number of points~\cite{kung1975finding}. Here, $n$ is the number of cascades, and while this numbers in the millions, this is a fast computation, as just described.

\section{Data Handling Costs}
\label{sec:costs}

An often overlooked part of image classification is the cost of loading and preparing the data prior to inference. Of the rare projects in the computer vision field that report inference speeds (e.g., the YOLO family of object detectors~\cite{redmon2016you,redmon2016yolo9000}), none report video decoding or image loading time. Likewise, the NoScope video query system~\cite{kang2017noscope} explicitly ignores these costs, claiming that GPU-based decoding is so fast as to be negligible or that decoding might be avoided altogether by obtaining raw video frames directly from the generating camera sensor.

This last point hints at why image loading or decoding costs \textit{should} be included when evaluating these systems: deployment scenarios for a given query system may differ drastically. Some may use top-of-the-line GPUs for video decoding, others may store video frames as individual image files on disk, while some may require video to be transported over a network prior to processing. Further, if multiple classification models with a variety of input representations might be used for classification, data handling costs---including preprocessing---must be included in throughput evaluations when deciding which classifier will be used at query time.

More concretely, for accurate model comparisons to be made, the throughput must be measured as the reciprocal of the average classification time, $t_\text{classify}$, defined as follows:
$$t_\text{classify} = t_\text{load} + t_\text{transform} + t_\text{infer}$$
where each time $t$ above is the average recorded over some typical set of input images, measured on the deployed system.

\begin{figure}
    \centering
    \includegraphics[width=.8\linewidth]{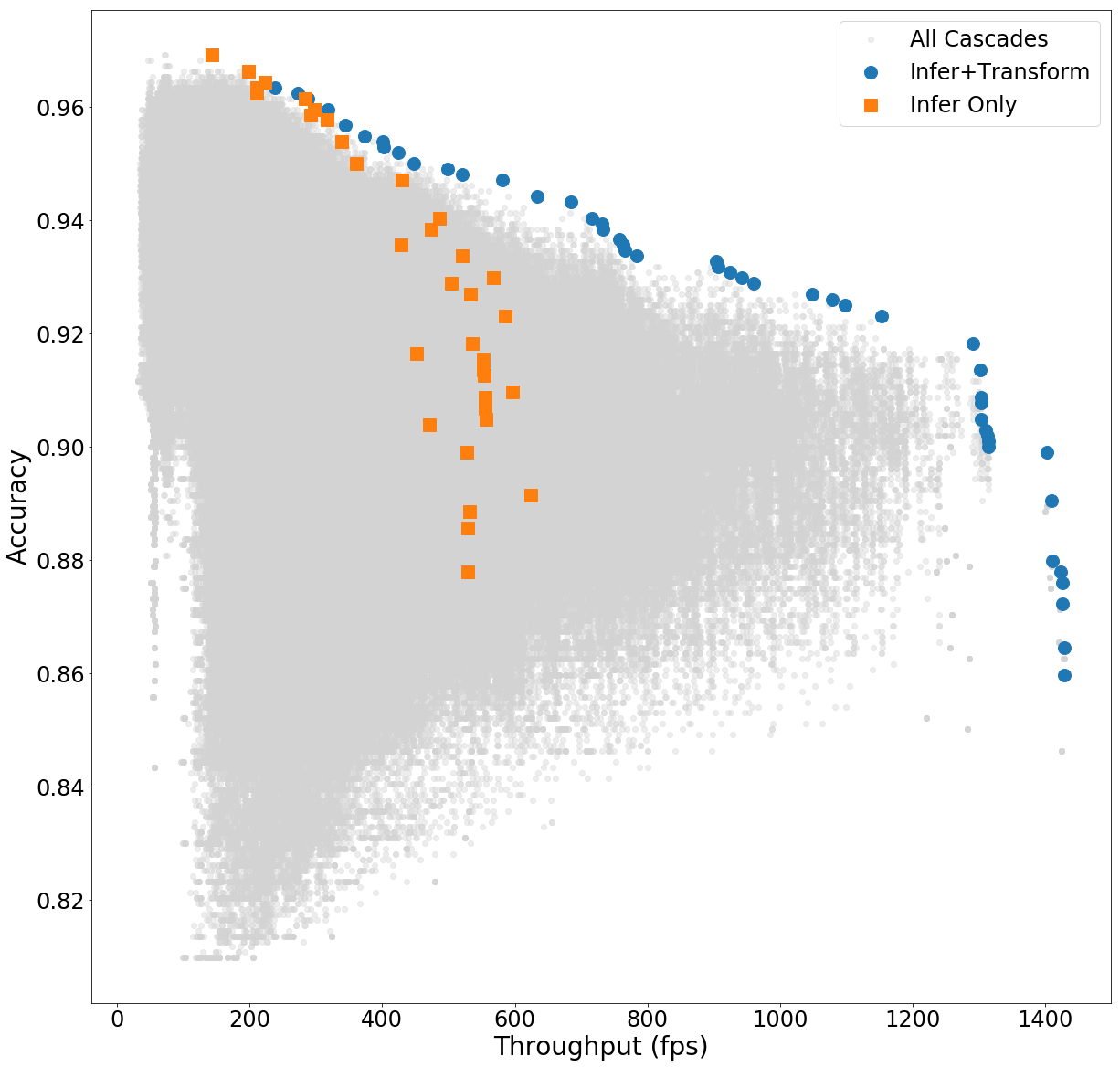}
    \caption{Cascades (gray) and Pareto frontier (blue) for an example deployment scenario, compared to the Pareto frontier for a  scenario only considering inference costs (orange). }
    \label{fig:paretoExample}
\end{figure}

Our experiments in Section~\ref{sec:expDataHandlingCosts} demonstrate that choosing among cascades with incorrect cost assumptions can lead to a large decrease in throughput. This throughput loss can be seen in Figure~\ref{fig:paretoExample}. The gray points depict all possible cascades for an example binary predicate (e.g., \texttt{contains-object(\textsf{\small semitruck})}) in a deployment scenario where image loading costs are negligible (full size raw data is already present in memory), but image preprocessing costs are incurred to transform the images into the appropriate resolutions and color representations for each model. The points on the Pareto frontier (shown in blue) represent the cascades that present the best tradeoffs between accuracy and throughput under this deployment scenario. The orange points show the cascades that would be on the Pareto frontier for this binary predicate if only inference costs were considered (pre-processed images are already present in memory). \textit{If the preprocessing costs were not considered, the query throughput would be far below its potential for most accuracy levels.}
 \section{Experiments}
\label{sec:experiments}

We have implemented \system\ as a prototype system, and in this section, we discuss our experiments that compare \system's performance against existing baseline methods, investigate the effects of data handling costs of different deployment scenarios, evaluate the effects of input transformations, and analyze the effects of increasing cascade depth.

\subsection{Experiment Setup}
\label{sec:experimentSetup}

To evaluate the methods and components used in \system, we designed a series of experiments with the following setup.

\minisection{Binary predicates}
We evaluated \system's performance over a set of 10 queries with a single \texttt{contains-object} binary predicate, randomly chosen from the 1,000 categories in the ImageNet dataset~\cite{ILSVRC15}. Each query is of the form ``SELECT * FROM images WHERE \texttt{contains-object}(\emph{category})'', where \emph{category} is one of those shown in Table~\ref{tab:queryNames}. The ImageNet dataset provides about 1,200 training images for each category. We held out 200 of each category's images as a validation set. The remaining images were labeled as positive examples. We then selected a matching number of random images drawn from the remaining 999 ImageNet categories and labeled them as negative examples. Because these training sets were relatively small, we  followed common data augmentation practice by creating a copy of each image that was flipped left-to-right, doubling the amount of training data.

We performed similar labeling for the validation set. For performance evaluation, we collected an additional dataset for each category using the image search functionality of the Google and Bing search engines, resulting in roughly 500 positive images per category. To find negative examples, we first created a large collection of images by performing image searches on Google and Bing for each of the remaining 999 ImageNet categories. From that, we randomly chose our negative examples to match the number of positive ones.

\minisection{Cascade configurations}
\label{sec:expCascadeConfigurations}
We used the Keras~\cite{chollet2015keras} deep learning library running TensorFlow\cite{tensorflow2015-whitepaper} to train and execute the CNNs used in our cascades. We varied several network architecture hyperparameters to create a range of models for each binary predicate: the number of convolutional layers (1, 2, 4), the number of convolutional nodes in each layer (16, 32), and the number of nodes in the final dense layer (16, 32, 64). These hyperparameters provide a range of reasonable options similar to those used in other systems (e.g.,~\cite{kang2017noscope}). We also varied the size of the input images (30x30, 60x60, 120x120, 224x224).  For each input size, we used five different image representations: full 3-channel color, each of the individual red, green, and blue color channels, and single-channel grayscale. In all, we constructed 360 simple CNNs for each binary predicate.

We also included in our pool of classifiers a fine-tuned implementation of ResNet50~\cite{he2016deep} that had been pre-trained on ImageNet. Fine-tuning was done using standard techniques; the final 1000-class classification layer was replaced with 64-node ReLu dense layer, followed by a 2-node softmax layer for the output of the binary prediction. These layers were trained using the same training set as the smaller, specialized classifiers.

For each classifier, we calibrated its decision thresholds using the validation dataset for five precision settings: 0.91, 0.93, 0.95, 0.97, and 0.99. We use each classifier variation to construct cascades of one and two levels, as well as three-level cascades with ResNet50 as the final layer (see Section~\ref{sec:cascadeDepth}), resulting in 1,301,405  possible cascades per predicate.

\begin{table}
    \centering
    \small
        \caption{ImageNet categories randomly selected for use as our experimental binary predicates.}
    \begin{tabular}{lclc}
    \toprule
           ~~~Predicate & ImageNet ID &   ~~~Query & ImageNet ID\\
    \midrule
      1.  \textsf{acorn} & n12267677 &  6.  \textsf{ferret} & n02443484\\
      2.  \textsf{amphibian} & n02704792 &  7. \textsf{komondor} & n02105505\\
      3.  \textsf{cloak} & n03045698 &  8.  \textsf{pinwheel} & n03944341\\
      4.  \textsf{coho} & n02536864 &  9.  \textsf{scorpion} & n01770393\\
      5.  \textsf{fence} & n03930313 &  10. \textsf{wallet} & n04548362\\

    \bottomrule
    \end{tabular}
    \label{tab:queryNames}
    \vspace{-.1cm}
\end{table}

\minisection{Deployment Scenarios}
\label{sec:costModels}
The data handling and preprocessing costs of particular deployment scenarios can have a large effect on the optimal choices in classifier cascades. To demonstrate this, we analyzed our cascades under four different scenarios, corresponding to those in Section~\ref{sec:designConsiderations}:

\begin{itemize}

    \item \cmNoCost\ --- This scenario ignores data handling and transformation costs---only inference costs (i.e., only the time required to evaluate the CNN) are considered when computing throughput, a practice commonly used in  computer vision literature.  However, as we show, the fastest inference often does not imply the fastest end-to-end query performance in practical deployments.

    \item \cmLoadResize\ --- This scenario includes the cost of loading a full-size image off an SSD hard disk, as well as the cost of resizing that image to the appropriate input size for a given classifier, as might be done when querying an existing corpus of archived image or video data.

    \item \cmLoad\ --- This scenario corresponds to deployments where images are resized on ingest before saving to disk. Load times are smaller, since the full-sized image is not loaded from disk if not needed for a particular classification. This scenario may occur when setting up a data collection system in tandem with a query system; proper image sizes for object detectors are known and initial transform costs can amortized over many queries.

    \item \cmResize\ --- This scenario only includes the computation costs of resizing the images, as in deployments where loading costs are negligible (e.g., images are loaded to memory directly from a connected camera sensor).

\end{itemize}

Data handling costs for the above scenarios only occur once for a given input: if a cascade includes two classifiers that use, for example, a 30x30 pixel red channel input, the costs to create that input are incurred only once per image.

\minisection{Cascade evaluation}
\label{sec:metrics}
To compare two sets of cascades in terms of throughput, we compute the area to the left of the curve (ALC) created by those points in this plot over a given accuracy range. Because a Pareto frontier is a collection of points and not a curve, we interpolate the curve as a step function.  Dividing the ALC  by the size of the accuracy range give the \emph{average throughput} for cascades in the Pareto frontier.
Dividing the ALC of one frontier by another gives the \emph{speedup} of the former over the latter. For fair comparisons, we use the accuracy range for the full set of cascades for each configuration and choose the smallest said range. In some cases, we compute the ALC for a Pareto frontier's cascades in a different cost context. These cascades are no longer a strict Pareto frontier, but we can still compute ALC for comparisons.
When comparing \system\ to a single classifier, such as against our ResNet50 baseline system, we choose the optimal cascade whose accuracy is both higher and closest to the accuracy of the single classifier.

\begin{figure}
    \centering
    \begin{tikzpicture}[trim axis left, trim axis right]
\begin{axis}[enlargelimits=false,
             width=.7\linewidth,
             height=.7\linewidth,
             axis lines*=left,
             axis on top,
             xlabel={Throughput (fps)},
             ylabel={Accuracy},
             legend cell align={left},
             legend style={font=\scriptsize, yshift=0.1cm, draw=none, anchor=north east, at={(1.05,1)}, legend columns=1, /tikz/every even column/.append style={column sep=0.1cm}},
             tickwidth=0,
                          label style={font=\scriptsize},
             xlabel style={yshift=0.1cm},
             ylabel style={xshift=-0.1cm},
             tick label style={font=\scriptsize}
             ]
        \addlegendimage{only marks, mark options={scale=0.5}, mark=*, color=gray,},
        \addlegendentry{All Cascades},
        \addlegendimage{only marks, mark options={scale=0.5}, mark=square*, color=red,},
        \addlegendentry{Baseline Cascades},
        \addlegendimage{only marks, mark options={scale=0.8}, mark=square*, color=red!50!black,},
        \addlegendentry{Baseline Optimal},
        \addlegendimage{only marks, mark options={scale=0.8}, mark=*, color=PYBLUE,},
        \addlegendentry{\system\ Optimal},

\addplot graphics [xmin=0,xmax=1500,ymin=0.6,ymax=0.99] {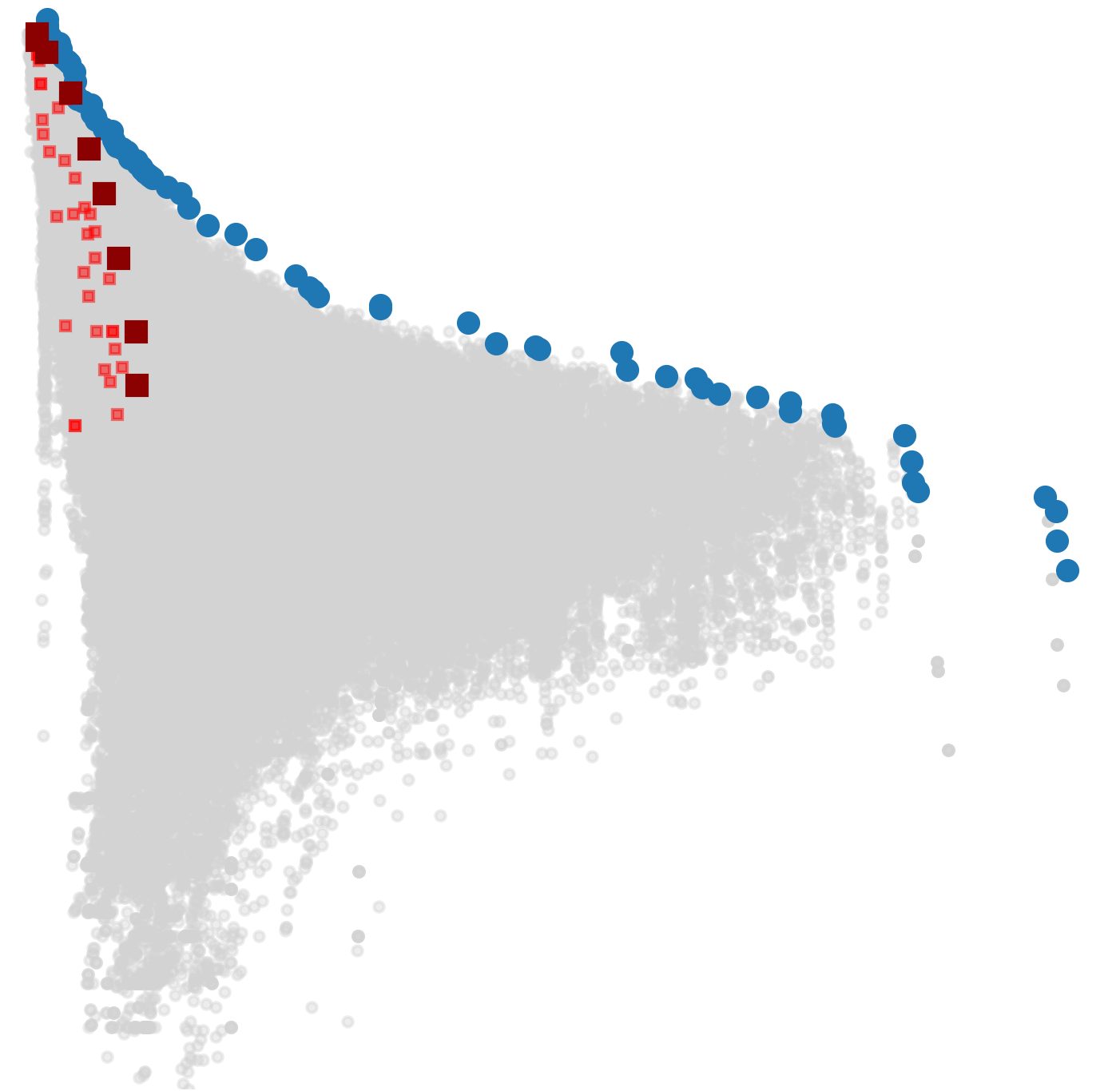};
\end{axis}
\end{tikzpicture}
\vspace{-.2cm}
    \caption{A comparison of the cascade space of \system\ (gray) compared to that of our Baseline cascades (red). \system's Pareto-optimal points are in blue. This example uses our \textsf{\small komondor} binary predicate under the \cmResize\ cost model. }
    \label{fig:noscopeCompare}
\vspace{-0.4cm}
\end{figure}

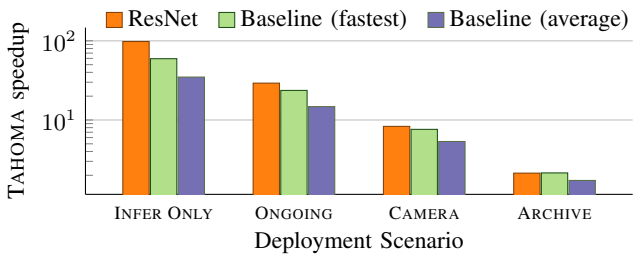
\begin{figure}
    \small
    \centering
            \pgfplotsset{
    compat=newest,
    every axis/.append style={
        legend image post style={xscale=.35}
    }
}
\begin{tikzpicture}

\pgfplotstableread{
X  Target FPS
1	resnet	98.101
2	noscopeMax	59.48
3	noscopeAvg	34.861
}\datatableNocost

\pgfplotstableread{
X  Target FPS
1	resnet	29.224
2	noscopeMax	23.64
3	noscopeAvg	14.746
}\datatableLoad

\pgfplotstableread{
X  Target FPS
1	resnet	8.305
2	noscopeMax	7.615
3	noscopeAvg	5.351
}\datatableResize

\pgfplotstableread{
X  Target FPS
1	resnet	2.131
2	noscopeMax	2.144
3	noscopeAvg	1.722
}\datatableLoadResize

\pgfplotstableread{
X  Target FPS
1	nocost	98.101
2	load	29.224
3	resize	8.305
4 loadresize	2.131
}\datatableResnet

\pgfplotstableread{
X  Target FPS
1	nocost	59.48
2	load	23.64
3	resize	7.615
4 loadresize	2.144
}\datatableNoScope

\pgfplotstableread{
X  Target FPS
1	nocost	34.861
2	load	14.746
3	resize	5.351
4 loadresize	1.722
}\datatableNoScopeFastest

\begin{axis}[
    ybar=0.5pt,
    axis lines*=left,
    ymajorgrids,
    /pgf/number format/1000 sep={,},
    bar width=0.35cm,
    enlarge x limits=0.2,
    width=\linewidth,
    height=3.8cm,
    legend style={
      font=\small,
      cells={anchor=west},
      legend columns=5,
      at={(0.52,0.92)},
      anchor=south,
      draw=none,
      fill=none,
      /tikz/every even column/.append style={column sep=0.2cm}
    },
        title style={font=\small, yshift=-.2cm,},
    xtick=data,
    xtick pos=left,
        xticklabels={\cmNoCost, \cmLoad, \cmResize, \cmLoadResize},
        xticklabel style={align=center, text height=1.5ex,font=\scriptsize},
        tickwidth=0,
        yticklabel style={/pgf/number format/fixed},
    legend=\empty,
    area legend,
        xlabel={Deployment Scenario},
    xlabel style={yshift=.05cm, align=center},
        ymin=0.0,
    ylabel={\system\ speedup},
    ylabel style={yshift=-.05cm, align=center, font=\small},
    ymode=log,
    ]

\addplot[plot2!50!black,fill=plot2] table [x=X, y=FPS] {\datatableResnet};
\addplot[plot3!50!black,fill=plot4] table [x=X, y=FPS] {\datatableNoScope};
\addplot[plot4!50!black,fill=plot5] table [x=X, y=FPS] {\datatableNoScopeFastest};

\addlegendentry{ResNet}
\addlegendentry{Baseline (fastest)}
\addlegendentry{Baseline (average)}

\end{axis}
\end{tikzpicture}
                                             \vspace{-.1cm}
    \caption{Average speedup values of \system\ over baselines. \textit{ResNet50} and \textit{Baseline (fastest)} comparisons use the optimal cascade with the nearest higher accuracy to ResNet50 and the fastest Baseline cascade, respectively. \textit{Baseline (average)} shows average speedups over the Baseline accuracy range. }
    \label{fig:baselines}
    
\end{figure}

\minisection{Hardware}
We used an Amazon EC2 p2.xlarge instance with an NVIDIA Tesla K80 GPU to do training, inference, and throughput measurements for all CNNs. For operations  suited to parallel CPU processing---such as finding the Pareto frontiers for our cascade sets---we used a 32-core (2.8GHz) Opteron 6100 server with 512GB RAM.

\subsection{Comparison Against Baselines}

\minisection{Overview}
We compare \system\ against a fine-tuned, pre-trained ResNet50 implementation~\cite{he2016deep}, as well as a set of non-optimized cascades, which comprise a subset of \system's design space. These two-level cascades  terminate in a full-cost classifier (i.e., ResNet50) and use full-color 224x224 images as input. These are similar to design to CNN-based cascades in previous work~\cite{kang2017noscope}. An illustration of the difference in the design spaces is shown in Figure~\ref{fig:noscopeCompare}, with \system's available cascade options being markedly larger due both the use of data transformations on the inputs and the additional cascade depth.

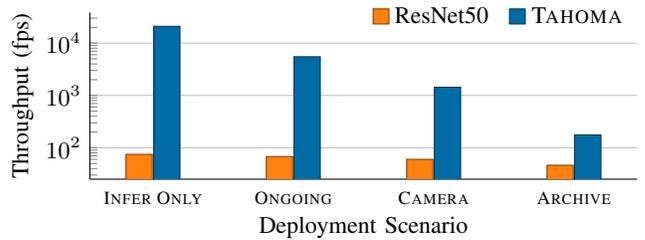
\begin{figure}
    \small
    \centering
                \pgfplotsset{
    compat=newest,
    every axis/.append style={
        legend image post style={xscale=.35}
    }
}
\begin{tikzpicture}

\pgfplotstableread{
X  Target FPS
1	nocost	74.8
2	load	67.8
3	resize	60
4	loadResize	46.2
}\datatableResnet

\pgfplotstableread{
X  Target FPS
1	nocost	20926
2	load	5484
3	resize	1433
4	loadResize	175.5
}\datatableSystem

\pgfplotstableread{
X  Target FPS
1	nocost	279.8
2	load	80.9
3	resize	23.9
4	loadResize	3.8
}\datatableSpeedup

\begin{axis}[
    ybar=0.5pt,
    axis lines*=left,
    ymajorgrids,
    /pgf/number format/1000 sep={,},
    bar width=0.35cm,
    enlarge x limits=0.15,
    width=\linewidth,
    height=3.8cm,
    legend style={
      font=\small,
      cells={anchor=west},
      legend columns=5,
      at={(0.75,0.85)},
      anchor=south,
      draw=none,
      fill=none,
      /tikz/every even column/.append style={column sep=0.2cm}
    },
        title style={font=\small, yshift=-.2cm,},
    xtick=data,
    xtick pos=left,
        xticklabels={\cmNoCost, \cmLoad, \cmResize, \cmLoadResize},
    xticklabel style={align=center, text height=1.5ex,font=\scriptsize},
        tickwidth=0,
        yticklabel style={/pgf/number format/fixed},
    legend=\empty,
    area legend,
        xlabel={Deployment Scenario},
    xlabel style={yshift=.05cm, align=center},
        ymin=0.0,
    ylabel={Throughput (fps)},
    ylabel style={yshift=-.05cm, align=center, font=\small},
    ymode=log,
    ]

\addplot[plot2!50!black,fill=plot2] table [x=X, y=FPS] {\datatableResnet};
\addplot[plot1!50!black,fill=plot1] table [x=X, y=FPS] {\datatableSystem};

\addlegendentry{ResNet50}
\addlegendentry{\system}

\end{axis}

\end{tikzpicture}
                                                             \vspace{-.2cm}
    \caption{Throughput of \system\ and  ResNet50 of fastest cascades for each cost model, averaged over 10 binary predicates. }
    \label{fig:maxFPS}
    \vspace{-.4cm}
\end{figure}

\begin{figure}[t]
    \small
    \centering
            \pgfplotsset{
    compat=newest,
    every axis/.append style={
        legend image post style={xscale=.35}
    }
}
\begin{tikzpicture}

\pgfplotstableread{
X  Target FPS
1	coral	3494
2	jackson 260
}\datatableNoscope

\pgfplotstableread{
X  Target FPS
1	coral	7153
2	jackson	7157
}\datatableTahoma

\pgfplotstableread{
X  Target FPS
1	coral	10700
2	jackson	7150
}\datatableTahomaDD

\begin{axis}[
    ybar=0.8pt,
    axis lines*=left,
    ymajorgrids,
    /pgf/number format/1000 sep={,},
    bar width=0.5cm,
    enlarge x limits=0.5,
    width=0.9\linewidth,
    height=3.8cm,
    legend style={
      font=\small,
      cells={anchor=west},
      legend columns=5,
      at={(0.7,0.94)},
      anchor=south,
      draw=none,
      fill=none,
      /tikz/every even column/.append style={column sep=0.2cm}
    },
        title style={font=\small, yshift=-.2cm,},
    xtick=data,
    xtick pos=left,
        xticklabels={\textsf{\scriptsize{coral}}, \textsf{\scriptsize{jackson}}},
        xticklabel style={align=center, text height=1.5ex,font=\scriptsize},
    ytick={0,2500,5000,7500,10000},
    tickwidth=0,
        scaled y ticks = false,
    y tick label style={/pgf/number format/fixed},
    legend=\empty,
    area legend,
            xlabel style={yshift=.05cm, align=center},
        ymin=0.0,
    ylabel={Throughput (fps)},
    ylabel style={yshift=-.05cm, align=center, font=\small},
        ]

\addplot[plot2!50!black,fill=plot2] table [x=X, y=FPS] {\datatableNoscope};
\addplot[plot4!50!black,fill=plot1] table [x=X, y=FPS] {\datatableTahomaDD};

\addlegendentry{NoScope}
\addlegendentry{\system+DD}

\end{axis}
\end{tikzpicture}
     \vspace{-.2cm}
    \caption{\new{Comparison with NoScope. \system+DD is  \system\ with a simulated NoScope-style difference detector. }}
        \label{fig:noscopeDirect}
    \vspace{-.2cm}

\end{figure}
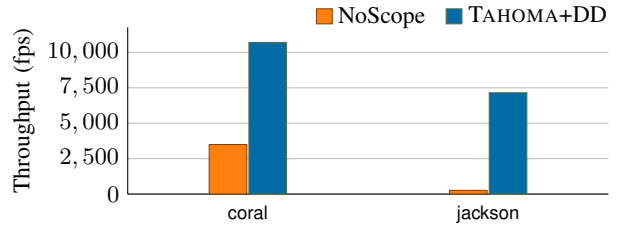

\minisection{Throughput gains}
Figure~\ref{fig:baselines} shows the performance of \system\ compared to our baselines for the four deployment scenarios from Section~\ref{sec:costModels}.  When only considering inference speed (i.e.,  \cmNoCost), \system\ yielded a 98x speedup over using a fine-tuned ResNet50 classifier alone, averaged over our 10 predicates.  \system\ showed a 35x average speedup over the accuracy range provided by the Baseline cascades. At the accuracy level provided by the fastest Baseline cascade for each predicate, \system\ yielded a 59x speedup on average.  For the other scenarios, data handling overheads reduce the speedup gains.  Nevertheless, \system\ achieves substantial speedup in all scenarios, and even the most expensive scenario that requires costly loading and transformation costs (\cmLoadResize) shows a nearly 2x speedup versus both ResNet50 and Baseline.

If speed is the priority, \system\ allows a user to trade accuracy for a large throughput boost. Figure~\ref{fig:maxFPS} shows \system's fastest optimal cascade compared to ResNet50. Across all predicates, the fastest cascades were not true cascades at all: they comprised a single specialized classifier with adequate accuracy and high throughput. In the \cmNoCost\ scenario, \system\ achieved an average throughput of 20,926 frames per second---280 times faster than our fine-tuned ResNet50 models, which had an average throughput of about 75 frames per second. The more realistic \cmLoad\ scenario still achieves 5484 frames per second---an 81x speedup. These large speedups come at a price, though: under \cmNoCost, the \system's fastest cascade was on average 12\% less accurate than ResNet50. Of course, as Figure~\ref{fig:noscopeCompare} shows, the optimal cascades provide a rich space of throughput and accuracy tradeoffs, so users can find the right balance for their needs.

\begin{figure*}
    \centering
    \begin{tikzpicture}
    \begin{groupplot}[
	group style={
		{horizontal sep=1.3cm, vertical sep=1cm, group size=4 by 1}
	},
    enlargelimits=false,
             width=.24\linewidth,
             height=.24\linewidth,
             axis lines*=left,
             axis on top,
             xlabel={Throughput (fps)},
             ylabel={Accuracy},
             legend cell align={left},
             legend style={fill=none,font=\small, draw=none, anchor=north east, at={(1.2,.92)}, legend columns=-1, /tikz/every even column/.append style={column sep=0.1cm}, yshift=1.3cm},
             tickwidth=0,
                         label style={font=\scriptsize},
             tick label style={font=\scriptsize},
             title style={font=\small, yshift=-.33cm}
             ]

\nextgroupplot [
    title={\textsf{amphibian}},
]

\addplot graphics [xmin=0,xmax=1500,ymin=0.45,ymax=0.97] {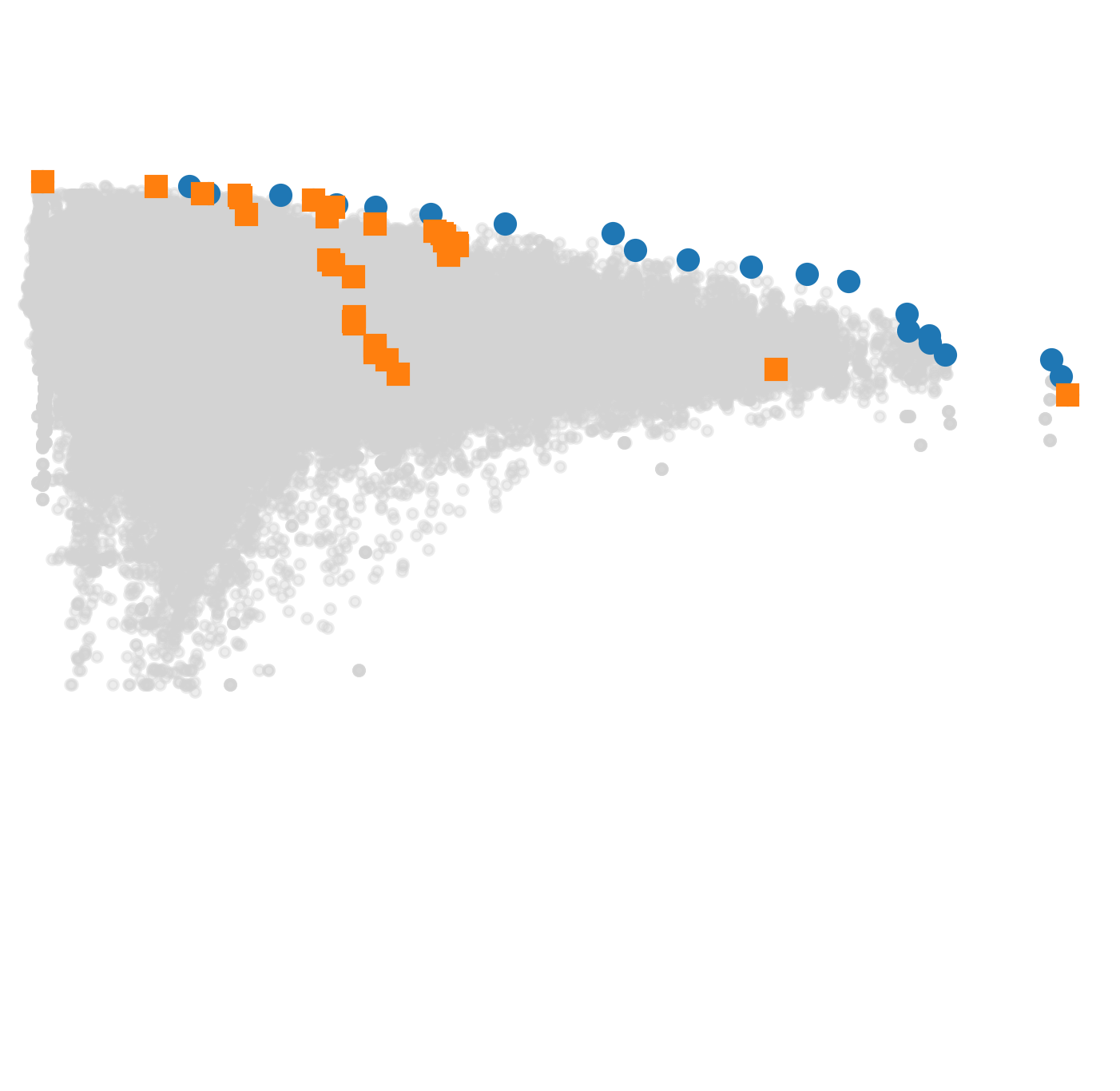};

\nextgroupplot [
    title={\textsf{fence}},
]
\addplot graphics [xmin=0,xmax=1500,ymin=0.45,ymax=0.97] {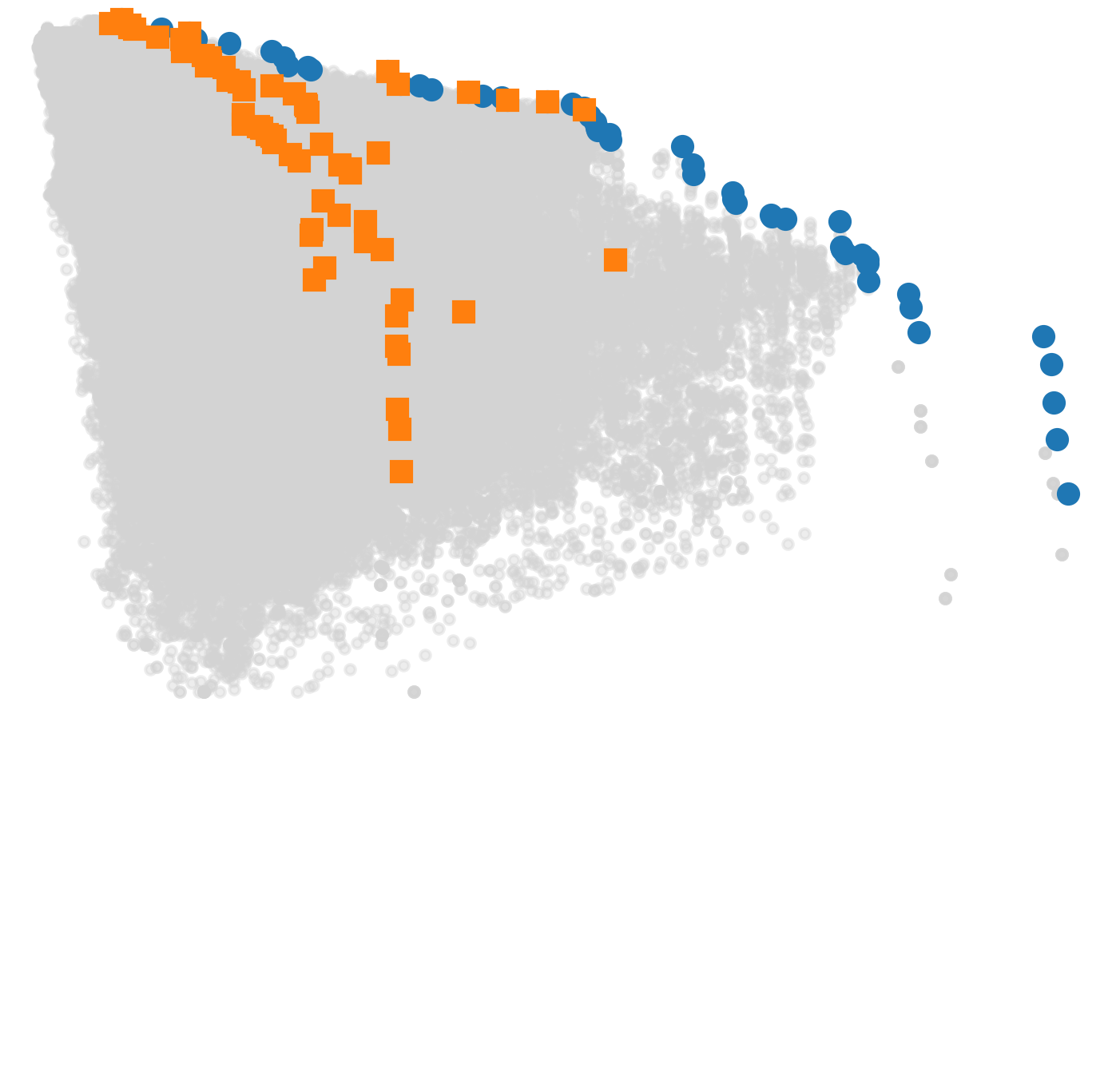};

\nextgroupplot [
    title={\textsf{scorpion}},
]

\addplot graphics [xmin=0,xmax=1500,ymin=0.45,ymax=0.97] {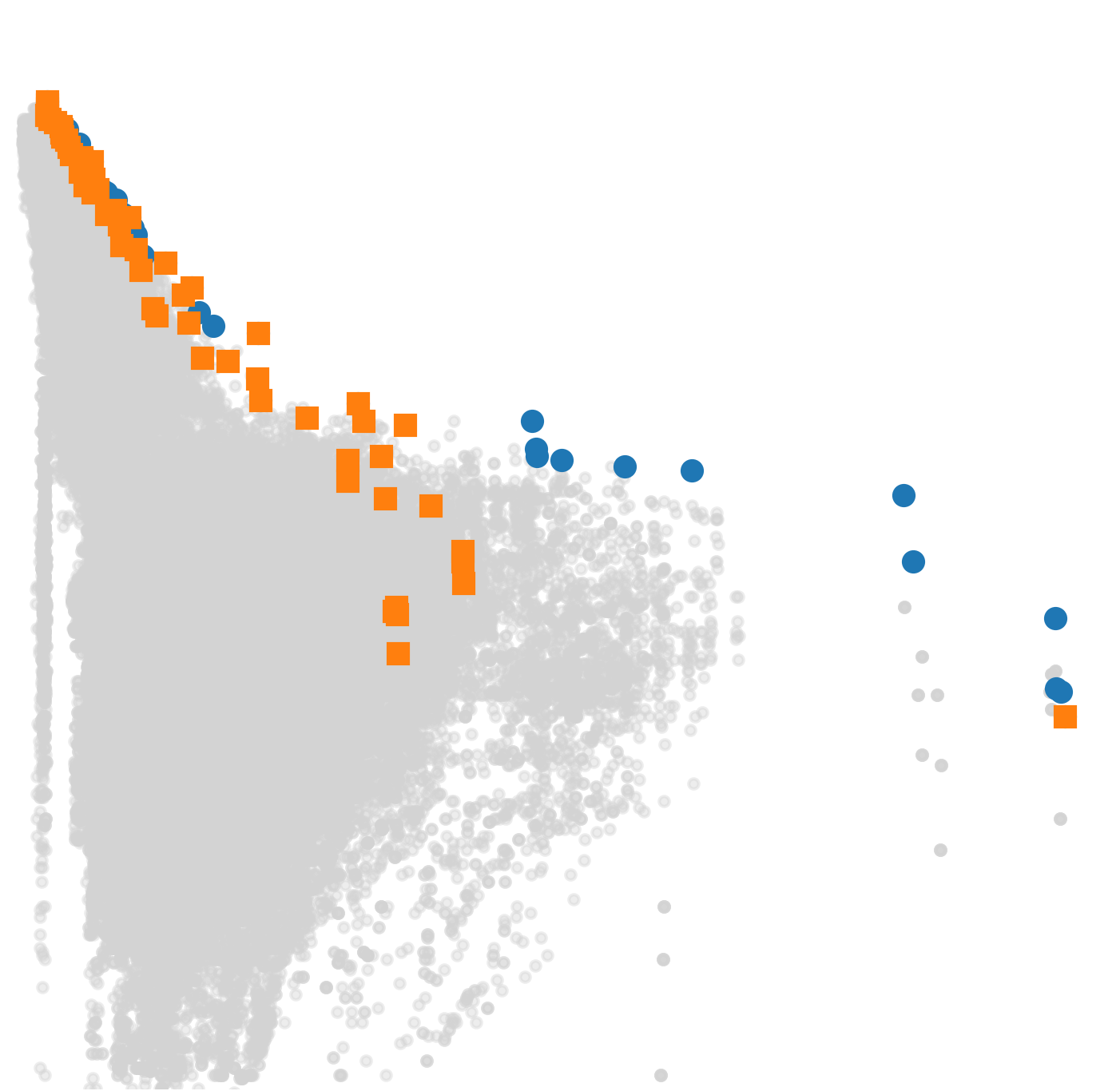};

\nextgroupplot [
    title={\textsf{wallet}},
]

\addlegendimage{only marks, mark options={scale=0.3}, mark=*, color=gray,}
\addlegendentry{All Cascades}
\addlegendimage{only marks, mark options={scale=0.5}, mark=*, color=PYBLUE,}
\addlegendentry{\cmResize}
\addlegendimage{only marks, mark options={scale=0.5}, mark=square*, color=orange,}
\addlegendentry{\cmNoCost}

\addplot graphics [xmin=0,xmax=1500,ymin=0.45,ymax=0.97] {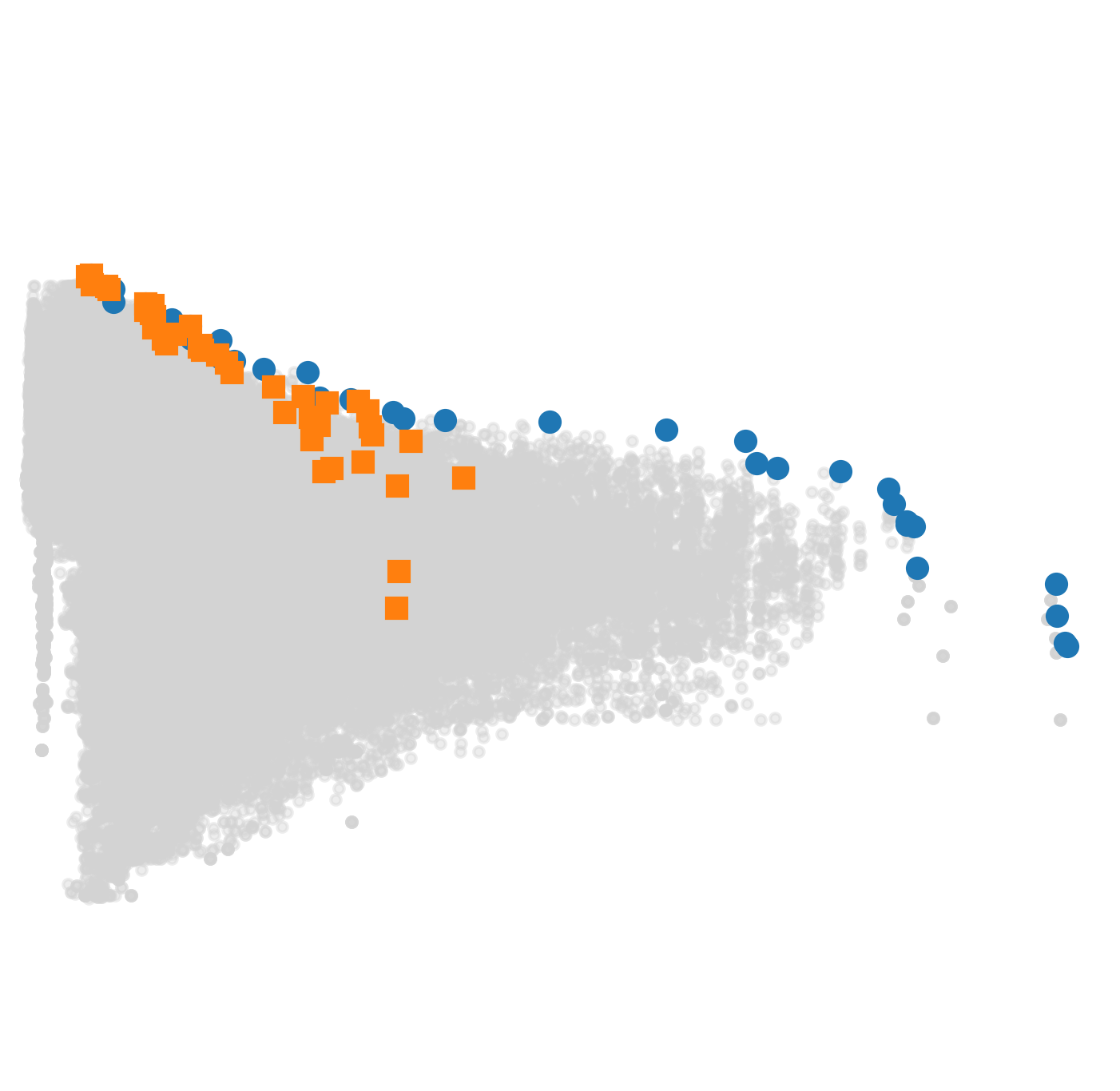};

\end{groupplot}
\end{tikzpicture}
    \caption{Pareto frontiers for several of our binary predicates, under the \cmResize\ deployment scenario (blue), compared to cascades that would be in the Pareto frontier for the \cmNoCost\ scenario (orange). Optimal cascades under one deployment scenario can significantly underperform those in another, so consideration of realistic deployment scenario costs is crucial.}
    \label{fig:resizePareto}
    \vspace{-0.3cm}
\end{figure*}

\subsection{Comparison with NoScope}
\label{sec:newNoscope}

\new{

Because NoScope~\cite{kang2017noscope} is the existing system most closely aligned with our work, we ran experiments to directly compare the two systems. For these experiments, we used the code and datasets (\textsf{\small coral} and \textsf{\small jackson}) provided by the NoScope authors\footnote{\url{https://github.com/stanford-futuredata/noscope}}. The other datasets presented in the NoScope paper were not publicly available. We used the default parameters provided in the NoScope code for each dataset and report results for both systems with a target precision of 0.95 used to select cascade thresholds. YOLOv2~\cite{redmon2016yolo9000} was used as the final, expensive classifier for both systems. Both NoScope and \system\ were run on AWS p2.xlarge instances. Note that CPU and GPU specifications differ between our NoScope installation and the one in the NoScope paper, so raw performance numbers differ between our experiments and theirs.

To compare NoScope and \system\ on an equal footing, we implemented \system+DD, which is \system\ with a simulated difference detector equivalent to that used by NoScope. The difference detector measures the similarity between the current frame and previously seen ones and reuses previous results if the compared frames meet a similarity threshold. This mechanism is orthogonal to our work and increase NoScope's throughput by avoiding many classifier executions. To create \system+DD, we recorded frame similarity using NoScope's difference detector and reused \system's results for frames meeting NoScope's threshold instead of classifying them.

Additionally, both systems used basic frame skipping, only processing one of every 30 frames. The results shown here include only those frames actively processed by each system, not those skipped this way. \system+DD results are measured in the \cmNoCost\ deployment scenario, which matches NoScope's throughput measurements. \system+DD results use the Pareto-optimal cascade with the closest but higher accuracy level to that of the NoScope for each dataset.

Figure~\ref{fig:noscopeDirect} compares the throughput for NoScope and \system+DD for the two public NoScope datasets. For both datasets, \system+DD significantly outperformed NoScope. On \textsf{\small coral}, \system+DD system reached a throughput of 10,700 fps, while NoScope's throughput was 3494 fps, giving \system+DD a 3.1x speedup over NoScope. On \textsf{\small jackson}, \system+DD system reached a throughput of 7,150 fps, while NoScope's throughput was 260 fps, giving \system+DD a 27.5x speedup over NoScope\footnote{
As may be apparent from the results, the \textsf{\scriptsize coral} dataset was a much simpler classification task than \textsf{\scriptsize jackson}, with far more reused results from the difference detector for \textsf{\scriptsize coral} (25.2\% reused) than for \textsf{\scriptsize jackson} (3.8\% reused). NoScope used the expensive YOLOv2 model for a significant number of frames on \textsf{jackson}, as well, leading to its slow performance. \system+DD's much larger cascade design space allowed it to find an accurate cascade that was able to avoid calling YOLOv2 for all but a few frames.
}.
}

\begin{table*}
    \small
    \centering
        \caption{Throughputs for various deployment scenarios when the cascade choices chosen in either oblivious or aware of scenario data handling costs. Here, permissible accuracy loss indicates how much accuracy the user is willing to trade for an increase in throughput. Scenario awareness can lead to significant throughput increases, shown in parentheses.}
    \begin{tabular}{lcccccc}
    \toprule
         \multirow{2}{*}[-2pt]{
         \begin{tabular}{l}Permissible\\accuracy loss\end{tabular}
         }   & \multicolumn{2}{c}{Scenario: \cmLoadResize} &  \multicolumn{2}{c}{Scenario: \cmResize} & \multicolumn{2}{c}{Scenario: \cmLoad}\\
         \cmidrule(lr){2-3}
         \cmidrule(lr){4-5}
         \cmidrule(lr){6-7}
          & ~~Oblivious~~ & Aware & ~~Oblivious~~ & Aware& ~~Oblivious~~ & Aware\\
    \midrule
        ~~~~\;~0\% loss  & 57.5 fps  & 58.3 fps (+1.4\%)
                         & 107.1 fps & 107.1 fps (+0.0\%)
                         & ~111.9 fps & ~111.9 fps (+0.0\%)\\
        ~~~~\;~2\% loss  & 85.1 fps  & 91.1 fps (+7.1\%)
                         & 267.5 fps & 324.6 fps (+21.3\%)
                         & ~985.2 fps & ~985.3 fps (+0.0\%)\\
        ~~~~\;~5\% loss  & 103.1 fps  & 117.1 fps (+13.5\%)
                         & 344.7 fps & 549.9 fps (+59.5\%)
                         & 1938.7 fps & 2000.8 fps (+3.2\%)\\
        ~~~~10\% loss    & 130.6 fps  & 142.0 fps (+8.7\%)~
                         & 568.0 fps & 806.8 fps (+42.0\%)
                         & 3669.1 fps & 3669.1 fps (+0.0\%)\\
    \bottomrule
    \end{tabular}
            \label{tab:costModelsAccLoss}
    \vspace{-.2cm}
\end{table*}

\subsection{Deployment Scenario Awareness}
\label{sec:expDataHandlingCosts}

\vspace{0.2cm}
Figure~\ref{fig:resizePareto} shows (in blue) the Pareto frontier of all classifier cascades for the \cmResize\ cost model for several of our binary predicates.  Additionally, the cascades that would be Pareto-optimal for each query under the \cmNoCost\ model are shown in orange. These orange points form a non-convex curve, since they are \emph{not a Pareto frontier} under the depicted \cmResize\ cost model. Each cascade may be impacted differently by loading and transformation costs, so their throughputs can change relative to one another in different deployment scenarios.
With few exceptions, the optimal cascades under \cmResize\ are different than the \cmNoCost\ ones. It is clear that if the data handling costs of a scenario like \cmResize\ were ignored and the ``optimal'' cascades were chosen only considering inference costs, considerable throughput gains would be missed.

Table~\ref{tab:costModelsAccLoss} shows the difference in throughput in  our deployment scenarios when cascades are chosen in a scenario-oblivious way (i.e., when only inference costs are considered, as in \cmNoCost) versus when the cascades are chosen taking scenario costs into consideration. Because \system\ provides a tradeoff between accuracy and throughput, we show results for four different levels of permissible accuracy loss. A user, for example, may decide that a 5\% decrease in accuracy is acceptable in order to process images faster. Then, in the \cmResize\ scenario, the system's throughput would increase by 59.5\% if cascades were chosen taking data handling costs into consideration, instead of being oblivious to these costs and only considering the classifier's inference.

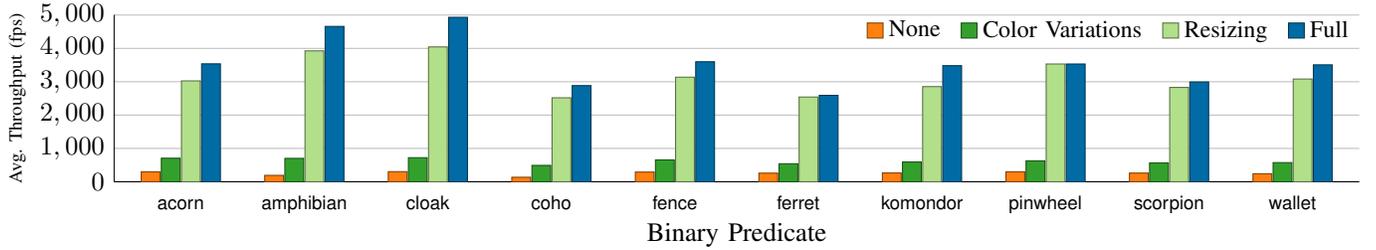
\begin{figure*}
    \centering
    \pgfplotsset{
    compat=newest,
    every axis/.append style={
        legend image post style={xscale=.35}
    }
}
\begin{tikzpicture}

\pgfplotstableread{
X  Target FPS
1	acorn	3536.73
2	amphibian	4658.939
3	cloak	4931.349
4	coho	2883.47
5	fence	3598.862
6	ferret	2591.469
7	komondor	3481.556
8	pinwheel	3531.734
9	scorpion	2992.858
10	wallet	3506.113
}\datatableFull

\pgfplotstableread{
X  Target FPS
1	acorn	298.251
2	amphibian	192.524
3	cloak	303.365
4	coho	135.412
5	fence	290.927
6	ferret	258.711
7	komondor	264.433
8	pinwheel	297.842
9	scorpion	261.255
10	wallet	240.348
}\datatableNone

\pgfplotstableread{
X  Target FPS
1	acorn	3025.055
2	amphibian	3925.842
3	cloak	4044.507
4	coho	2518.339
5	fence	3136.257
6	ferret	2541.39
7	komondor	2853.56
8	pinwheel	3531.734
9	scorpion	2829.766
10	wallet	3080.731
}\datatableResizing

\pgfplotstableread{
X  Target FPS
1	acorn	706.657
2	amphibian	701.775
3	cloak	722.207
4	coho	491.358
5	fence	655.469
6	ferret	540.837
7	komondor	593.735
8	pinwheel	627.111
9	scorpion	564.975
10	wallet	574.334
}\datatableColors

\begin{axis}[
    ybar=0.5pt,
    axis lines*=left,
    ymajorgrids,
    /pgf/number format/1000 sep={,},
    bar width=0.25cm,
    enlarge x limits=0.06,
    width=\textwidth,
    height=3.8cm,
    legend style={
      font=\small,
      cells={anchor=west},
      legend columns=5,
      at={(0.8,0.78)},
      anchor=south,
      draw=none,
      fill=none,
      /tikz/every even column/.append style={column sep=0.2cm}
    },
        title style={font=\small, yshift=-.2cm,},
    xtick=data,
    xtick pos=left,
        xticklabels={\textsf{acorn},\textsf{amphibian},\textsf{cloak},\textsf{coho},\textsf{fence},\textsf{ferret},\textsf{komondor},\textsf{pinwheel},\textsf{scorpion},\textsf{wallet}},
    xticklabel style={font=\scriptsize, align=center, text height=1.5ex},     ytick={0,1000,2000,3000,4000,5000},
    tickwidth=0,
        yticklabel style={/pgf/number format/fixed},
    legend=\empty,
    area legend,
        xlabel={Binary Predicate},
    xlabel style={yshift=.1cm, align=center},
    ymax=5000,
    ymin=0.0,
    ylabel={Avg. Throughput (fps)},
    ylabel style={yshift=-.05cm, align=center, font=\scriptsize},
    ]

\addplot[plot2!50!black,fill=plot2] table [x=X, y=FPS] {\datatableNone};
\addplot[plot3!50!black,fill=plot3] table [x=X, y=FPS] {\datatableColors};
\addplot[plot4!50!black,fill=plot4] table [x=X, y=FPS] {\datatableResizing};
\addplot[plot1!50!black,fill=plot1] table [x=X, y=FPS] {\datatableFull};

\addlegendentry{None}
\addlegendentry{Color Variations}
\addlegendentry{Resizing}
\addlegendentry{Full}

\end{axis}
\end{tikzpicture}
     \vspace{-0.6cm}
    \caption{Average throughput of optimal cascades for cascade sets that use different input transformations.}
    \label{fig:dataTransformFPS}
    \vspace{-.4cm}
\end{figure*}

\begin{figure}
    \centering
    \pgfplotsset{scaled x ticks=false}
\pgfplotsset{
legend image code/.code={
\draw[mark repeat=2,mark phase=2]
plot coordinates {
(0cm,0cm)
(0.18cm,0cm)        (0.36cm,0cm)         };}
}

\begin{tikzpicture}
    \tikzset{
      every pin/.style={color=black,font=\tiny,pin distance=.15cm,inner sep=1pt,},
	      }
\begin{groupplot}[
	group style={
		{horizontal sep=.7cm, vertical sep=1cm, group size=1 by 1}
	},
	enlarge y limits=0.06,
	enlarge x limits=0.03,
	width=0.65\linewidth,
	height=0.65\linewidth,
				ymax=1,
	xmin=0,
	xmax=1500,
		tick label style={font=\scriptsize},
		tickwidth=0.1cm,
    xtick pos=left,
	ytick pos=left,
                	xlabel style={yshift=0.1cm,font=\small},
	ylabel style={yshift=-0.1cm,font=\small},
	title style={yshift=-0.2cm,font=\small},
	scaled y ticks=false,
  xlabel={Throughput (fps)},
  ylabel={Accuracy}
]

\nextgroupplot [
	xlabel={Throughput (fps)},
	legend columns=1,
  legend cell align={left},
	legend style={
      fill=none,
      font=\scriptsize,
	    at={(1.05,0.0)},
	    anchor=south west,
	    /tikz/every odd column/.append style={column sep=0.05cm},
	    /tikz/every even column/.append style={column sep=0.5cm},
        draw=none
    }
]

\addplot[color=plot2, mark=diamond*, mark options={draw=plot2!50!black, scale=1}] coordinates {
(59.99, 0.98363)
(272.77, 0.96503)
(398.44, 0.95908)
(417.14, 0.95610)
(814.80, 0.95015)
(829.91, 0.89732)
(1218.34, 0.89286)
(1222.08, 0.88765)
(1232.42, 0.87872)
(1401.01, 0.87723)
(1411.62, 0.86682)
(1414.87, 0.85268)
(1419.13, 0.83929)
(1434.25, 0.81920)};

\addplot[color=plot1, mark=square*, mark options={draw=plot1!50!black, scale=.8}] coordinates {
(160.74, 0.99330)
(181.06, 0.98958)
(191.80, 0.98810)
(221.35, 0.98735)
(259.28, 0.98661)
(271.24, 0.98214)
(289.82, 0.97917)
(356.46, 0.97768)
(426.70, 0.97396)
(470.98, 0.97098)
(579.27, 0.96652)
(626.75, 0.96503)
(678.98, 0.96429)
(732.80, 0.96354)
(766.36, 0.96280)
(782.35, 0.96131)
(789.45, 0.95833)
(797.67, 0.95610)
(799.22, 0.95387)
(800.06, 0.95312)
(816.70, 0.95164)
(817.33, 0.94940)
(914.37, 0.94717)
(929.03, 0.94048)
(929.55, 0.93676)
(982.50, 0.93006)
(982.95, 0.92783)
(986.95, 0.92634)
(1034.24, 0.92188)
(1034.50, 0.92113)
(1126.82, 0.91964)
(1128.37, 0.90997)
(1129.26, 0.90923)
(1134.19, 0.90774)
(1156.37, 0.90699)
(1163.78, 0.90551)
(1164.06, 0.90402)
(1164.61, 0.89732)
(1218.34, 0.89286)
(1222.08, 0.88765)
(1232.42, 0.87872)
(1401.01, 0.87723)
(1411.62, 0.86682)
(1414.87, 0.85268)
(1419.13, 0.83929)
(1434.25, 0.81920)
};

\addplot[color=plot3, mark=*, mark options={draw=plot3!50!black, scale=.7}] coordinates {
(160.74, 0.99330)
(181.06, 0.98958)
(191.80, 0.98810)
(221.35, 0.98735)
(259.28, 0.98661)
(271.24, 0.98214)
(289.82, 0.97917)
(356.46, 0.97768)
(426.70, 0.97396)
(470.98, 0.97098)
(579.27, 0.96652)
(626.75, 0.96503)
(678.98, 0.96429)
(732.80, 0.96354)
(766.36, 0.96280)
(782.35, 0.96131)
(789.45, 0.95833)
(797.67, 0.95610)
(799.22, 0.95387)
(800.06, 0.95312)
(816.70, 0.95164)
(817.33, 0.94940)
(914.37, 0.94717)
(929.03, 0.94048)
(929.55, 0.93676)
(982.50, 0.93006)
(982.95, 0.92783)
(986.95, 0.92634)
(1034.24, 0.92188)
(1034.50, 0.92113)
(1126.82, 0.91964)
(1128.37, 0.90997)
(1129.26, 0.90923)
(1134.19, 0.90774)
(1156.37, 0.90699)
(1163.78, 0.90551)
(1164.06, 0.90402)
(1164.61, 0.89732)
(1218.34, 0.89286)
(1222.08, 0.88765)
(1232.42, 0.87872)
(1401.01, 0.87723)
(1411.62, 0.86682)
(1414.87, 0.85268)
(1419.13, 0.83929)
(1434.25, 0.81920)
};

\addplot[color=plot5, mark=triangle*, mark options={draw=plot5!50!black, scale=1.1}] coordinates {
(160.59, 0.99405)
(160.74, 0.99330)
(167.51, 0.99256)
(170.39, 0.99182)
(178.29, 0.99107)
(213.13, 0.99033)
(250.70, 0.98884)
(259.28, 0.98661)
(304.34, 0.98512)
(361.76, 0.98214)
(377.50, 0.97991)
(383.68, 0.97693)
(410.20, 0.97619)
(414.92, 0.97545)
(517.95, 0.97470)
(535.46, 0.97024)
(658.86, 0.96949)
(682.01, 0.96875)
(705.73, 0.96726)
(721.55, 0.96577)
(725.02, 0.96503)
(763.39, 0.96429)
(764.22, 0.96354)
(772.97, 0.96280)
(778.35, 0.96205)
(782.35, 0.96131)
(783.74, 0.96057)
(785.71, 0.95982)
(790.73, 0.95908)
(792.77, 0.95759)
(792.87, 0.95685)
(797.67, 0.95610)
(821.46, 0.95461)
(888.05, 0.95238)
(897.01, 0.95089)
(897.04, 0.94940)
(902.65, 0.94866)
(902.83, 0.94792)
(914.37, 0.94717)
(929.03, 0.94048)
(929.55, 0.93676)
(941.41, 0.93527)
(950.93, 0.93378)
(960.93, 0.93304)
(982.50, 0.93006)
(982.95, 0.92783)
(1002.91, 0.92634)
(1014.35, 0.92560)
(1014.37, 0.92485)
(1023.32, 0.92411)
(1031.88, 0.92336)
(1066.42, 0.92262)
(1123.49, 0.92039)
(1126.82, 0.91964)
(1130.43, 0.91220)
(1134.14, 0.91071)
(1134.25, 0.90774)
(1156.37, 0.90699)
(1163.78, 0.90551)
(1164.06, 0.90402)
(1164.61, 0.89732)
(1218.34, 0.89286)
(1222.08, 0.88765)
(1232.42, 0.87872)
(1401.01, 0.87723)
(1411.62, 0.86682)
(1414.87, 0.85268)
(1419.13, 0.83929)
(1434.25, 0.81920)
};

\addplot[color=plot4, mark=pentagon*, mark options={draw=plot4!50!black, scale=.8}] coordinates {
(160.59, 0.99405)
(160.74, 0.99330)
(167.51, 0.99256)
(170.39, 0.99182)
(178.29, 0.99107)
(213.13, 0.99033)
(250.70, 0.98884)
(259.28, 0.98661)
(304.34, 0.98512)
(361.76, 0.98214)
(377.50, 0.97991)
(383.68, 0.97693)
(410.20, 0.97619)
(414.92, 0.97545)
(517.95, 0.97470)
(535.46, 0.97024)
(658.86, 0.96949)
(682.01, 0.96875)
(705.73, 0.96726)
(721.55, 0.96577)
(725.02, 0.96503)
(763.39, 0.96429)
(764.22, 0.96354)
(772.97, 0.96280)
(778.35, 0.96205)
(782.35, 0.96131)
(783.74, 0.96057)
(785.71, 0.95982)
(790.73, 0.95908)
(792.77, 0.95759)
(792.87, 0.95685)
(797.67, 0.95610)
(821.46, 0.95461)
(888.05, 0.95238)
(897.01, 0.95089)
(897.04, 0.94940)
(902.65, 0.94866)
(902.83, 0.94792)
(914.37, 0.94717)
(929.03, 0.94048)
(929.55, 0.93676)
(941.41, 0.93527)
(950.93, 0.93378)
(960.93, 0.93304)
(982.50, 0.93006)
(982.95, 0.92783)
(1002.91, 0.92634)
(1014.35, 0.92560)
(1014.37, 0.92485)
(1023.32, 0.92411)
(1031.88, 0.92336)
(1066.42, 0.92262)
(1123.49, 0.92039)
(1126.82, 0.91964)
(1130.43, 0.91220)
(1134.14, 0.91071)
(1134.25, 0.90774)
(1156.37, 0.90699)
(1163.78, 0.90551)
(1164.06, 0.90402)
(1164.61, 0.89732)
(1218.34, 0.89286)
(1222.08, 0.88765)
(1232.42, 0.87872)
(1401.01, 0.87723)
(1411.62, 0.86682)
(1414.87, 0.85268)
(1419.13, 0.83929)
(1434.25, 0.81920)
};

\addplot[color=plotGray, mark=star, mark options={draw=plotGray!50!black, scale=1.0}] coordinates {
(153.55, 0.99479)
(160.59, 0.99405)
(167.44, 0.99330)
(167.51, 0.99256)
(175.20, 0.99182)
(206.88, 0.99107)
(245.31, 0.99033)
(253.73, 0.98884)
(254.50, 0.98735)
(261.47, 0.98661)
(282.85, 0.98586)
(304.34, 0.98512)
(307.21, 0.98438)
(328.47, 0.98363)
(353.35, 0.98289)
(373.65, 0.98214)
(386.14, 0.98140)
(387.18, 0.98065)
(401.20, 0.97991)
(407.62, 0.97842)
(416.08, 0.97768)
(416.32, 0.97693)
(440.20, 0.97619)
(517.95, 0.97470)
(531.51, 0.97396)
(537.15, 0.97321)
(618.79, 0.97247)
(635.60, 0.97098)
(638.85, 0.97024)
(658.86, 0.96949)
(682.01, 0.96875)
(705.73, 0.96726)
(721.55, 0.96577)
(740.99, 0.96503)
(763.39, 0.96429)
(764.22, 0.96354)
(772.97, 0.96280)
(778.35, 0.96205)
(782.35, 0.96131)
(783.74, 0.96057)
(785.71, 0.95982)
(790.73, 0.95908)
(792.77, 0.95759)
(792.87, 0.95685)
(797.67, 0.95610)
(821.46, 0.95461)
(850.58, 0.95312)
(888.05, 0.95238)
(897.01, 0.95089)
(897.04, 0.94940)
(902.65, 0.94866)
(902.83, 0.94792)
(914.37, 0.94717)
(929.03, 0.94048)
(929.55, 0.93676)
(941.41, 0.93527)
(950.93, 0.93378)
(960.93, 0.93304)
(982.50, 0.93006)
(982.95, 0.92783)
(1002.91, 0.92634)
(1014.35, 0.92560)
(1014.37, 0.92485)
(1023.32, 0.92411)
(1031.88, 0.92336)
(1066.42, 0.92262)
(1123.49, 0.92039)
(1126.82, 0.91964)
(1130.43, 0.91220)
(1134.14, 0.91071)
(1134.25, 0.90774)
(1156.37, 0.90699)
(1163.78, 0.90551)
(1164.06, 0.90402)
(1164.61, 0.89732)
(1218.34, 0.89286)
(1222.08, 0.88765)
(1232.42, 0.87872)
(1401.01, 0.87723)
(1411.62, 0.86682)
(1414.87, 0.85268)
(1419.13, 0.83929)
(1434.25, 0.81920)
};

\addlegendentry{1 level}
\addlegendentry{1 level + ResNet}
\addlegendentry{2 level}
\addlegendentry{2 level + ResNet}
\addlegendentry{3 level}
\addlegendentry{3 level + ResNet}

\end{groupplot}
\end{tikzpicture}
     \vspace{-0.2cm}
    \caption{Evolution of Pareto frontier as cascades depth increases, shown for our \textsf{\small fence} predicate in the \cmResize\ scenario. Other predicates and scenarios showed similar results.  As cascades get deeper, Pareto frontier improvements become negligible.}
    \label{fig:cascadeDepth}
\end{figure}
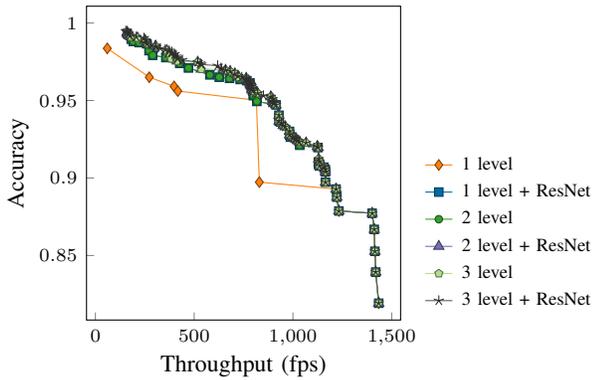

\subsection{Analysis of Input Transformations}

\vspace{0.2cm}
\system\ uses several different input transformations to expand the space of simple classifiers used to construct cascades. To see how these affect \system's performance, we constructed four cascade sets that used  varying subsets of the transformations: \textit{None}, which used no input transformations (i.e., all inputs are 224x224 three-color-channel images); \textit{Color Variations}, which used only the transformations that extract the color channels or create grayscale images; \textit{Resizing}, which used only transformations that reduce resolution; and \text{Full}, which used the full set of transforms included in \system.

Figure~\ref{fig:dataTransformFPS} shows the average throughput for each cascade set for each binary predicate,  computed using the ALC method described in Section~\ref{sec:metrics}. We computed these values over the accuracy range of the \textit{Full} cascade set for each predicate. Image resizing operations by far have the largest impact on throughput, giving nearly a ten-fold increase over \textit{None}. The resized 3-channel, 30x30 pixel input images equate to 2,700 input values, while the full-sized 3-channel, 224x224 pixel images equate to 150,528 values; this huge reduction in input size results in orders of magnitude fewer tensor operations during CNN inference. Likewise, reducing the color depth of an image from three channels to one reduces the CNN's computational requirement by two thirds.  These transforms, especially resolution reduction, are critical to high query throughput; they enable much smaller CNNs, more than paying for the transforms' computational costs.

\subsection{Analysis of Increased Cascade Depth}
\label{sec:cascadeDepth}

\vspace{0.2cm}
Each additional level added to the cascade exponentially increases the size of the cascade design space. Evaluating the 1.3 million cascades used in results reported elsewhere in this section is fast, taking about 1 minute on average per binary predicate. However, this cascade set includes a full cross-product only for one- and two-level cascades, with three-level cascades restricted to those with fine-tuned ResNet50 as the final classifier. If we considered all possible three-level cascades from all available models, our cascade set balloons to about 45 million distinct cascades, requiring about 40 minutes of evaluation time per predicate. Evaluating a full cross-product of four-level cascades is intractable (roughly $360^4$ total cascades).

Figure~\ref{fig:cascadeDepth} shows how a cascade set's Pareto frontier evolves as the maximum depths of the cascades increase. Each set of cascades includes all depths up to its maximum.  A ``one level'' cascade simply corresponds to our set of basic classification models. Where a cascade depth indicates ``+ ResNet50'', we append ResNet50 as an additional final level of the cascade. That is, "Two level + ResNet" cascades comprise two levels populated by any model from our collection, followed by a final ResNet50 level. The addition of a ResNet50 level effectively doubles the number of cascades in a cascade set. Adding a full additional level (drawn from our full set of models) increases the cascade set size and evaluation time exponentially.

As can be seen, increasing the depth of the cascades has diminishing returns to throughput at query time, while greatly increasing computational cost of cascade evaluation during system initialization. Moving from ``Two level + ResNet50'' to ``Three level (full)'' only increases average throughput by 1.0\%, while increasing evaluation time nearly 40-fold to just over 40 minutes. Thus, for our experiments, we have restricted cascades to at most ``Two level + ResNet50''.The minimal increase in throughput capabilities of additional layers is not worth the huge increase in evaluation time.

 \section{Related Work}
\label{sec:relatedWork}

\system\ builds upon a rich history of work in the database, image processing, and machine learning communities.

\minisection{Query systems for visual data}
A number of  database and query systems targeted at visual data have been developed over the past two decades~\cite{hu2011survey,snoek2008concept}. Early approaches  involved manual textual labeling  (\eg~\cite{smith1992stratification}) or extracting rudimentary low-level features (\eg~\cite{flickner1995query,ren2009state}). Later systems performed additional semantic object extraction, using hand-written functions and statistical methods to define objects~\cite{petkovic1999content, petkovic2000framework}.
That work, however, precedes the recent deep learning revolution in computer vision and relies on highly manual object extraction methods that limit system applicability and scale.

Work that extended relational query methods over visual data~\cite{li1997moql,kuo1996content,golshani1998language} may be useful in extending query capabilities in a system like ours. Fagin's work with the Garlic system~\cite{fagin1998fuzzy}, for example, deals with a number of issues particular to querying multimedia databases and the fuzziness of data extracted from multimedia data. However, the data handled by \system\ does not exhibit the same kind of fuzziness as Garlic, which is concerned with approximate matches and image similarity. Our system deals with binary predicates that---while based on probabilistic classifier output---are considered strictly true or false at query time.

There has been a recent explosion of interest in query systems for visual data due to advances in classifier accuracy made possible by convolutional neural networks and to the massive datasets made possible by cheap storage and image sensors. NoScope~\cite{kang2017noscope}, for example, uses several techniques, including classifier cascades, to accelerate query processing over video.
Our key departure from this prior work is to include transformations of the input image representation within our query plans, drastically expanding the design space of classifier cascades and enabling much smaller models and order-of-magnitude throughput improvements. We compare our our system with NoScope in Section~\ref{sec:newNoscope}.

\new{BlazeIt~\cite{kang2018blazeit} optimizes queries for objects found in video, in part by using small, specialized CNNs to quickly answer queries where possible. Like NoScope, these specialized CNNs do not make use of Tahoma's input transformations. The rich space of Tahoma's specialized CNNs, however, could potentially be integrated into a query optimizer like BlazeIt.
Focus~\cite{hsieh2018focus} is a system that indexes objects in live video. It too, uses specialized CNNs to speed up queries. Focus varies the resolution of input images when creating the set of specialized CNNs, though it does not use \system's other input transformations. Additionally, \system's use of cascades of multiple specialized CNNs (rather than a single CNN) creates a design space of millions of possible specialized cascades and a much richer set of Pareto-optimal choices. Further, \system\ can quickly evaluate its cascades in deployment-specific settings, determining the cascades optimal for the deployment's current operating characteristics. This could be particularly useful in dynamic scenarios, such as with networked cameras with varying bandwidth constraints.} VideoStorm~\cite{zhang2017live} is a system that automatically adjusts parameters like resolution and frame rate to maximize output quality of computer vision algorithms, based on computing resource availability in large clusters.

\minisection{Image classification}
Deep CNNs have revolutionized the field of computer vision, leading to breakthroughs in image classification and detection capabilities in recent years. The ImageNet competition~\cite{ILSVRC15} has had as one of its core challenges a 1000-category, million image classification task. A deep CNN was first used in 2012 and greatly reduced the best error rate~\cite{hinton2015distilling}, and the error rate has dropped in the subsequent years to near or better than human performance on the same task (e.g,~\cite{ioffe2015batch,he2016deep,hu2017squeeze}).  Deep CNNs can now facilitate semantic content extraction from images and videos, supporting the development of large scale visual analytics databases. Deep CNNs are too slow, however, in their current incarnations to be applied at scale. Research in speeding up CNNs has seen some recent interest (e.g., SqueezeNet~\cite{squeezenet} and MobileNets~\cite{howard2017mobilenets}).  \system\ is essentially classifier-agnostic, so these and future networks can be incorporated into our cascading techniques.

\minisection{Classifier cascades}
Our work leans heavily on previous research into classifier cascades. One of the first works to use the technique was the Viola-Jones face detector~\cite{viola2004robust}, which used a series of classifiers based on simple image features to detect faces in photographs. More recently, cascades have been used to accelerate the slow inference speeds of deep neural networks~\cite{kang2017noscope,cai2015learning,sun2013deep}. Other work has used cascades to improve accuracy, rather than speed~\cite{verschae2008unified}. Chen et al.~\cite{chen2012classifier} used classifier cascades to reduce the cost of feature extraction for text. Our system, rather, uses feature extraction (in the form of input transformations) in cascades to create smaller, faster models. We show how classifier cascade-driven query optimization can be exploited in a visual data analytics system.

\minisection{Model selection and management}
\system's general method of generating a large number of model (and cascade) variations and selecting among those can be seen as a form of model selection~\cite{kumar2016model}. A number of recent data systems have been proposed or developed to assist in machine learning model creation, management, and deployment~\cite{feng2012towards,crankshaw2014missing,bailis2017macrobase,kraska2013mlbase}. These systems have been developed with general machine learning tasks in mind, while \system\ focuses directly on performing queries and analytics over visual data. Because of this tight focus, we can take advantage of characteristics of CNN architectures and properties of image and video data. Our model generation and selection methods could be integrated with systems like Velox~\cite{crankshaw2014missing} or MacroBase~\cite{bailis2017macrobase}, whose functionality in serving models and handling large amounts of fast data would be complementary to our work.
 
\section{Conclusion and Future Work}
\label{sec:conclusion}

In this paper, we have presented a method of accelerating content extraction from large corpora of visual data, with the aim of supporting visual analytics queries. We showed how constructing a huge number of classifier cascades from a wide variety of CNN-based classification models can yield large speedups in content extraction. Our cascades applied input transformations on the raw image corpus to further reduce classification costs. We also demonstrated the necessity of being deployment scenario-aware---that is, considering costs such as those for data loading and image transformation---when evaluating the accuracy and throughput tradeoffs of classifiers.

While this paper primarily focused on image classification tasks, it is just the beginning of series of work that will develop a  data analytics for visual data that will take full advantage of spatio-temoporal locality present in adjacent video frames to further accelerate content extraction. We hope to include new state-of-the-art computer vision methods to extract more complex data, which will then allow the processing of complex analytical queries over video.

\section*{Acknowledgment}
This work was supported by grants from the Toyota Research Institute and a University of Michigan MIDAS grant.
 
\bibliographystyle{abbrv}
\bibliography{ms}

\balance
\end{document}